\documentclass{aa}

\usepackage{graphicx}
\usepackage{multirow}
\usepackage{txfonts}
\usepackage[colorlinks=true]{hyperref}
\begin{document}

   \title{Plausibility of ultraviolet burst generation in the low solar chromosphere}

   \author{Lei Ni\inst{1,2,3,4}
          \and
          Guanchong Cheng\inst{1,4}
          \and
         Jun Lin\inst{1,3,4} 
          }

   \institute{Yunnan Observatories, Chinese Academy of Sciences, Kunming, Yunnan 650216, P. R. China \\
              \email{leini@ynao.ac.cn}
        \and
              Key Laboratory of Solar Activity, National Astronomical of Observatories, Chinese Academy of Sciences, Beijing 100012, P. R. China.  
        \and 
              Center for Astronomical Mega-Science, Chinese Academy of Sciences, 20A Datun Road, Chaoyang District, Beijing 100012, P. R. China        
         \and
              University of Chinese Academy of Sciences, Beijing 100049, P. R. China.
         }

 %  \date{Received September 15, 1996; accepted March 16, 1997}

\abstract{
\emph{Context.} Ultraviolet (UV) bursts and Ellerman bombs (EBs) are small-scale magnetic reconnection events taking place in the highly stratified, low solar atmosphere. The plasma density, reconnection mechanisms, and radiative cooling and transfer processes clearly differ from one layer of the atmosphere to the next. In particular, EBs are believed to form in the upper photosphere or the low chromosphere. It is still not clear whether UV bursts have to be generated at a higher atmospheric layer than the EBs or whether both UV bursts and EBs can occur in the low chromosphere.   
  
\emph{Aims.} We numerically studied the low $\beta$ magnetic reconnection process around the solar temperature minimum region (TMR) by including more realistic physical diffusions and radiative cooling models. We aim to find out whether UV bursts may occur in the low chromosphere and to investigate the dominant mechanism that accounts for heating in the UV burst in the chromosphere.
  
\emph{Methods.} We used the single-fluid magnetohydrodynamic (MHD) code NIRVANA to perform the simulations. The time-dependent ionization degrees of hydrogen and helium are included in the code, which lead to a more realistic magnetic diffusion caused by electron-neutral collision and ambipolar diffusion. A more realistic radiative cooling model from \cite{Carlsson2012} is included in the simulations. The initial mass density and temperature are $1.66057\times10^{-6}$\,kg\,m$^{-3}$ and $4400$\,K, respectively, values that are typical for the plasma environment around TMR.   
  
\emph{Results.} Our results in high resolution indicate that the plasmas in the reconnection region are heated up to more than $20,000$\,K if the reconnecting magnetic field is as strong as $500$\,G, which suggests that UV bursts can be generated in the dense low chromosphere. The dominant mechanism for producing the UV burst in the low chromosphere is heating, as a result of the local compression in the reconnection process. The thermal energy occurring in the reconnection region rapidly increases after the turbulent reconnection mediated by plasmoids is invoked. The average power density of the generated thermal energy in the reconnection region can reach over $1000$\,erg\,cm$^{-3}$\,s$^{-1}$, which is comparable to the average power density accounting for a UV burst.  With the strength of the reconnecting magnetic field exceeding $900$\,G, the width of the synthesized Si IV 1394 {\AA} line profile with multiple peaks can reach up to $100$\,km\,s$^{-1}$, which is consistent with observations.
}

\keywords{  Magnetic reconnection; Magnetohydrodynamics (MHD); Heating mechanisms; Radiative cooling; Sun: chromosphere
               }
\maketitle
 
 \titlerunning {Plausibility of ultraviolet burst generation in the low solar chromosphere} 
 
%
%--------------------------------------------------------------------------------------------------------------------------------%

\section{Introduction}\label{sec:intro}

Partially ionized plasma exists in many astrophysical environments. How the interaction among the neutrals and ionized plasmas that affects magnetic reconnection is still an open question. The low solar atmosphere is naturally a good laboratory for providing suitable opportunities to study magnetic reconnection in partially ionized plasma. Numerous small-scale events of reconnection in the low solar atmosphere have been observed in multi-wavelengths by advanced solar telescopes of high resolution and the fine structures of these events have also been recognized \citep[e.g.,][]{Peter2014, Xue2016, Tian2018, Huang2018, Romano2019, da Silva Santos2020, Yan2020, Song2020, Joshi2021, Rast2021, Hou2021}.

One of the most important discoveries by the \emph{Interface Region Imaging Spectrograph} \citep[IRIS;][]{DePontieu2014} is the UV burst \citep[e.g.,][]{Peter2014, Tian2016, Grubecka2016, Young2018, Vissers2019, Chen2019a, Chen2019b}, which is a small, intense transient brightening seen in ultraviolet images. Observational results indicate that the local heating in UV bursts is mainly caused by magnetic reconnection in the low solar atmosphere. Such UV bursts have strong emissions in Si~{\sc{iv}} 1400 \AA\  and are also bright in 1600 \AA\  and 1700 \AA\ . A strong emission in Si~{\sc{iv}} 1400 \AA\  requires a temperature $\sim 2\times 10^4$\,K below the middle chromosphere or $ \sim 8 \times 10^4$\,K in the upper chromosphere \citep{Rutten2016}. The plasma density in the low chromosphere is about two to five orders of magnitude higher than those above the middle chromosphere \citep{Avrett2008}. The broad discrepancies in terms of the plasma environments can result in different magnetic reconnection mechanisms at different chromospheric layers \citep{Ni2020, Ni2018b, Jara-Almonte2019, Jara-Almonte2021}.    

The 3D Radiation Magnetohydrodynamics (RMHD) simulations carried out by \cite{Hansteen2019} show that the current sheet with UV emissions is essentially located above the middle chromosphere, which is also supported by the recent radiative hydrodynamic simulations by revisiting the spectral features of the UV bursts \citep{Hong2021}. We point out, on the other hand, that the reconnection process that accounts for the UV burst, in the simulation of \cite{Hansteen2019}, was driven by the numerical diffusion. Furthermore, the non-LTE (non-local thermal equilibrium) inversion of the Swedish 1-m Solar Telescope \citep[SST;][]{Scharmer2003} and the IRIS data \citep{Vissers2019} indicates that the high temperature (up to $2\times 10^{4}$~K) events could appear around TMR; compared with the results of LTE inversion, the emission in Si~IV could be seen at a lower altitude.

In the past few years, several teams have focused on studies of magnetic reconnection mechanisms in the partially ionized plasma in the low solar atmosphere \citep[e.g.,][]{Sakai2008, Leake2012, Leake2013, Murphy2015, Ni2015, Ni2018a, Alvarez2017, Singh2019, Peter2019, Jara-Almonte2019}. For the first time, 2.5 D single-fluid MHD simulations at high resolution which include physical diffusions have shown that the plasma in the low chromosphere can be heated above tens of thousands of Kelvin in a low $\beta$ magnetic reconnection process \citep{Ni2015}. However, the time-independent ionization degree has resulted in an underestimation of the radiative cooling from the reconnection region in previous works \citep{Ni2015, Ni2016, Ni2021} and it also leads to overestimate of magnetic diffusion due to the electron-neutral collision, as well as of the ambipolar diffusion due to the separation of ions from neutrals. Subsequent reactive multi-fluid MHD simulations that included the interaction among ions and neutrals proved that the non-equilibrium ionization-recombination makes the temperature increase more difficult, but the plasma around TMR can still be heated to a temperature above $20,000$\,K when the reconnection magnetic field is stronger than 500\,G \citep{Ni2018a, Ni2018b}. Such a strong magnetic field of several hundred to several thousand Gauss in the low solar atmosphere is usually observed or derived from magnetic field extrapolations in active regions \citep[e.g.,][]{Yan2017, Leenaarts2018, Getling2019, Wang2020}. 

In order to investigate if the temperature in the low chromosphere can be heated up to above $20,000$\,K by magnetic reconnection to form UV bursts, more accurate physical diffusivities as well as a better radiative cooling model are needed. Previous multi-fluid MHD simulations in the low chromosphere have indicated that the ionized plasma and neutrals are well coupled and no significant difference exists between the ion temperature and neutral temperature in the low $\beta$ reconnection process \citep{Ni2018a}. Therefore, it is still reasonable to use the single-fluid MHD simulation to study low $\beta$ reconnection processes such as UV bursts.

In this work, we optimize the one-fluid MHD code NIRVANA \citep[version 3.8;][]{Ziegler2011} and we consider the time-dependent ionization degree. We include both hydrogen and helium in the simulation to get more accurate ambipolar diffusion and magnetic diffusion caused by the electron-neutral collision. Two different radiative cooling models are applied to studying the effect of radiative cooling on magnetic reconnection. The time-dependent ionization degrees make the radiative cooling process and the related physical diffusivities more realistic than previous one-fluid MHD simulations \citep{Ni2015,Ni2016,Ni2021}. The rest of the paper is structured as follows. The model and numerical approach are described in Section 2. Our results and the associated discussions are presented in Section. 3. We offer our summary and outlook for the future in Section 4.  

%--------------------------------------------------------------------------------------------------------------------------------%
\section{Numerical setup} \label{sec:model}

\subsection{MHD equations and important coefficients} \label{sec:equations}

In this work, we assume that the plasmas are composed with hydrogen atoms, helium atoms, electrons, and ions contributed by ionized hydrogen and helium. All the components are considered as one fluid, and the decoupling of ions and neutrals are embodied by the ambipolar diffusion effect. The optimized single-fluid MHD code NIRVANA is applied to implement the 2.5D MHD simulations. The solved MHD equations are as follows:

\begin{eqnarray}
\frac{\partial \rho }{\partial t}&=&-\nabla \cdot \left (\rho \mathbf{v}\right ) \\
\frac{\partial \left ( \rho \mathbf{v} \right )}{\partial t} &=&- \nabla \cdot \left [ \rho \mathbf{vv} +\left ( p+\frac{1}{2\mu _{0}}\left | \mathbf{B} \right |^{2} \right )I-\frac{1}{\mu _{0}}\mathbf{BB} \right ] \nonumber \\
&&+\nabla \cdot \tau  \\
\frac{\partial e}{\partial t}&=&-\nabla \cdot\left [ \left ( e+p+\frac{1}{2\mu _{0}}\left | \mathbf{B} \right |^{2} \right )\mathbf{v} \right ]  \nonumber \\
&&+\nabla \cdot\left [ \frac{1}{\mu _{0}}\left ( \mathbf{v}\cdot \mathbf{B} \right )\mathbf{B}   \right ] \nonumber \\
&&+\nabla \cdot \left [ \mathbf{v} \cdot \tau_S +\frac{\eta }{\mu _{0}}\mathbf{B}\times \left ( \nabla \times \mathbf{B} \right )\right ]       \nonumber \\
&&-\nabla \cdot \left [ \frac{1 }{\mu _{0}}\mathbf{B}\times\mathbf{E}_{AD}\right ] \nonumber \\
&&+Q_{rad}+H  \label{energy} \\
\frac{\partial \mathbf{B}}{\partial t} &=& \nabla \times \left ( \mathbf{v}\times \mathbf{B}-\eta \nabla \times \mathbf{B}+\mathbf{E}_{AD} \right )  \label{indeq} 
\end{eqnarray}
\begin{eqnarray}
e&=&\frac{p}{\gamma -1}+\frac{1}{2}\rho \left | \mathbf{v} \right |^{2}+\frac{1}{2\mu _{0}}\left | \mathbf{B} \right |^{2} \\
p&=&\frac{\left (1.1+Y_{iH}+0.1Y_{iHe} \right )\rho}{1.4m_{i}}k_{B}T
\end{eqnarray}
where $\rho$, $\mathbf{v}, \mathbf{B}, p, T, e, Y_{iH}, Y_{iHe}$ are the mass density, fluid velocity, magnetic field, thermal pressure, temperature, total energy density, and ionization fractions of hydrogen and helium, respectively. The number density of the total helium is assumed to be $10\%$ of the total hydrogen, and we only consider the primary ionization of helium. Also, $m_{i}$ is the mass of proton, and $k_B$ is the Boltzmann constant. We set the ratio of specific heats as $\gamma = 5/3$. The stress tensor is $\tau_S=\xi \left [ \nabla \mathbf{v}+\left ( \nabla \mathbf{v} \right )^{\mathrm{T}}-\frac{2}{3}\left ( \nabla\cdot  \mathbf{v}  \right )I \right ]$, where $\xi$ is the dynamic viscosity coefficient and its unit is kg m$^{-1}$ s$^{-1}$.  Since the current sheet we study in this work is assumed to be parallel to the solar surface, we ignore the gravity and the initial plasma density is taken as a constant. The maximum temperature in our simulation is about tens of thousands of Kelvin, the heat conduction does not play a significant role \citep{Ni2021} and can be ignored in this work. Functions $Q_{rad}$ and $H$ refer to radiative cooling and heating in the energy equation, which will be described in greater detail in this paper.

We include the physical magnetic diffusion in our simulations \citep{Khomenko2012}, which is given by:
\begin{equation}
\eta=\eta_{ei}+\eta_{en}=\frac{m_{e} \nu_{ei} }{e_c^{2}n_{e}\mu_{0}}+\frac{m_{e} \nu_{en} }{e_c^{2}n_{e}\mu_{0}}
\end{equation}
where $m_e$ is the mass of the electron, $e_c$ is the electron charge,  $\mu_0$ is the magnetic permeability coefficient in vacuum, $n_e$ is the electron density calculated as $n_e=\rho (Y_{iH}+0.1Y_{iHe})/(1.4m_i)$, $\nu_{ei}$ and $\nu_{en}$ are frequencies of electron-ion and electron-neutral collisions, respectively, which are determined more accurate here than before \citep{Ni2015, Ni2016} such that:

\begin{equation}
\nu_{ei} = \frac{n_e e_c^4 \Lambda}{3m_e^2\epsilon_0^2} \left( \frac{m_e}{2\pi k_B T} \right)^{3/2},
\label{nuei}
\end{equation}

\begin{equation}\
\nu_{en} = n_n \sqrt{\frac{8k_B T}{\pi m_{en}}} \sigma_{en}
\label{nuen}
\end{equation}
where $\epsilon_0$ is the permittivity of vacuum, $\Lambda$ is the Coulomb logarithm, $n_n$ is the number density of the neutral particles, and $\sigma_{en}$ is the collision cross section. $m_{en}=m_e m_n/(m_e+m_n)$; we can obtain $m_{en}\simeq m_e$ because the mass of the neutral particle $m_n$ is much greater than the electron mass $m_e$. The expression of $\Lambda$ is given by:
\begin{equation}
\Lambda = 23.4-1.15\log_{10}n_e +3.45\log_{10}T
\label{nuei}
\end{equation}
with $n_e$ expressed in cgs units and $T$ in eV. Since we include the helium, the collision frequency $\nu_{en}$ is contributed by collisions between electrons and neutral hydrogen and collisions between electrons and neutral helium, respectively. Therefore, $\nu_{en}$ can be written into the following formulae:
\begin{equation}
\nu_{en} = n_{nHe} \sqrt{\frac{8k_B T}{\pi m_{e}}} \sigma_{e-nHe} + n_{nH} \sqrt{\frac{8k_B T}{\pi m_{e}}} \sigma_{e-nH}
\label{nuen}
\end{equation}
where $n_{nHe}=0.1\rho (1-Y_{iHe})/(1.4m_i)$ is the number density of the neutral helium and  $n_{nH}=\rho (1-Y_{iH})/(1.4m_i)$ is the number density of the neutral hydrogen. Also, $\sigma_{e-nHe}$ and $\sigma_{e-nH}$ are the cross sections for electron-neutral helium collision and electron-neutral hydrogen collision, respectively. We take $\sigma_{e-nH}=2\times10^{-19}$\,m$^2$ and $\sigma_{e-nHe}=\sigma_{e-nH}/\sqrt{3}$ according to \cite{Vranjes2013}. Then, the magnetic diffusions can be calculated as:
\begin{equation}
\eta_{ei}\simeq 1.0246\times10^8 \Lambda T^{-1.5},
\label{etaei}
\end{equation}
\begin{equation}
\eta_{en}\simeq 0.0351\sqrt{T}\frac{\left[\frac{0.1}{3}(1-Y_{iHe})+(1-Y_{iH})\right]}{Y_{iH}+0.1Y_{iHe}}.
\label{etaen}
\end{equation}
The unit for $\eta_{ei}$ and $\eta_{en}$ in Eq. (\ref{etaei}) and Eq. (\ref{etaen}) is m$^2$\,s$^{-1}$.

The ambipolar diffusion electric field $\mathbf{E}_{AD}$ in the energy Eq. (\ref{energy}) and induction Eq. (\ref{indeq}) is given by:
\begin{equation}
\mathbf{E}_{AD}=\frac{1}{\mu_{0}}\eta _{AD}\left [ \left ( \nabla \times \mathbf{B} \right ) \times \mathbf{B}  \right ] \times \mathbf{B} 
\end{equation}
where $\eta_{AD}$ is the ambipolar diffusion coefficient.  The formula of $\eta_{AD}$ is as below \citep{Khomenko2012, Ni2020}:
\begin{equation}
\eta_{AD}=\frac{(\rho_n/\rho)^2}{\rho_i \nu_{in}+\rho_e \nu_{en}},
\label{etaAD}
\end{equation}
 the unit of $\eta_{AD}$ is m$^3$\,s\,kg$^{-1}$. Since both the hydrogen and helium are included, we can get:
\begin{equation}
 \rho_n/\rho = \frac{0.4(1-Y_{iHe})+(1-Y_{iH})}{1.4},
\end{equation} 

\begin{eqnarray}
 \rho_i \nu_{in} &=& \rho_{iH} n_{nH} \sqrt{\frac{8k_BT}{\pi m_i/2}} \sigma_{iH-nH}+ \nonumber  \\
                           && \rho_{iH} n_{nHe} \sqrt{\frac{8k_BT}{\pi 4m_i/5}} \sigma_{iH-nHe}+        \nonumber \\
                            && \rho_{iHe}n_{nH}\sqrt{\frac{8k_BT}{\pi 4 m_i/5}} \sigma_{iHe-nH}+ \nonumber \\
                            &&  \rho_{iHe} n_{nHe} \sqrt{\frac{8k_BT}{\pi 2m_i}} \sigma_{iHe-nHe},                          
\end{eqnarray}
 where $\rho_{iH}=\rho Y_{iH}/1.4$ is the ionized hydrogen density, $\rho_{iHe}= 0.4\rho Y_{iHe}/1.4$ is the ionized helium density, and $\sigma_{iH-nH}$, $\sigma_{iH-nHe}$, $\sigma_{iHe-nH}$, and $\sigma_{iHe-nHe}$ are the cross sections for ionized hydrogen-neutral hydrogen collision, ionized hydrogen-neutral helium collision, ionized helium-neutral hydrogen collision and ionized helium-neutral helium collision, respectively. We take $\sigma_{iH-nH} = 1.5\times10^{-18}$\,m$^2$, $\sigma_{iH-nHe} = \sigma_{iHe-nH} =\sigma_{iHe-nHe}= \sigma_{iH-nH}/\sqrt{3}$ based on previous calculations of integral elastic cross sections \citep{Barata2010, Vranjes2013}. The electron collision part is written as:
 \begin{eqnarray}
  \rho_e \nu_{en} &=& \rho_e n_{nH} \sqrt{\frac{8k_BT}{\pi m_e}} \sigma_{e-nH} +  \nonumber  \\
                            &   & \rho_e n_{nHe} \sqrt{\frac{8k_BT}{\pi m_e}} \sigma_{e-nHe}. \label{rhonuen}
\end{eqnarray}
 Since $\rho_e/\sqrt{m_e} << \rho_{iH}/\sqrt{m_i}$ and $\sigma_{e-nHe} < \sigma_{e-nH}<\sigma_{iH-nH}$, we ignore the electron collision in Eq. (\ref{rhonuen}), and the ambipolar diffusion coefficient is simplified as $\eta_{AD}\simeq (\rho_n/\rho)^2/ (\rho_i \nu_{in})$.
 
Since neither the Saha ionization equilibrium nor the coronal equilibrium are valid in the chromosphere, we use the temperature dependent ionization degree based on the RADYN test atmosphere results by solving the radiative transfer equations \citep{Carlsson2012}. The temperature-dependent $Y_{iH}$ and $Y_{iHe}$ are shown in Figure 1. We point out that the ionization degree of hydrogen from the tests shows a spread around the fitted curves \citep{Carlsson2012}. Therefore, the actual magnetic diffusion caused by electron-neutral collision and ambipolar diffusion could be overestimated or underestimated when we follow the curve of $Y_{iH}$ in Figure1. The curve of $Y_{iHe}$ (as shown in Figure1) is a good fit down to $50\%$ ionization at $T=10^4$\,K,  but it underestimates the amount of HeI from there to $T=3.5\times10^4$\,K, which causes the magnetic diffusion due to electron-neutral collision and ambipolar diffusion to be underestimated in this temperature range. 

However, we note that the neutral number densities of hydrogen and helium always decrease with temperature; this trend will not change and the diffusions caused by neutrals will no longer dominate when the plasma is heated to a high temperature to generate UV emissions no matter whether the spread exists or not. Therefore, the spread around the fitted curves does not change the main results and conclusions of this work. 
\begin{figure}
\centering
\includegraphics[width=\hsize]{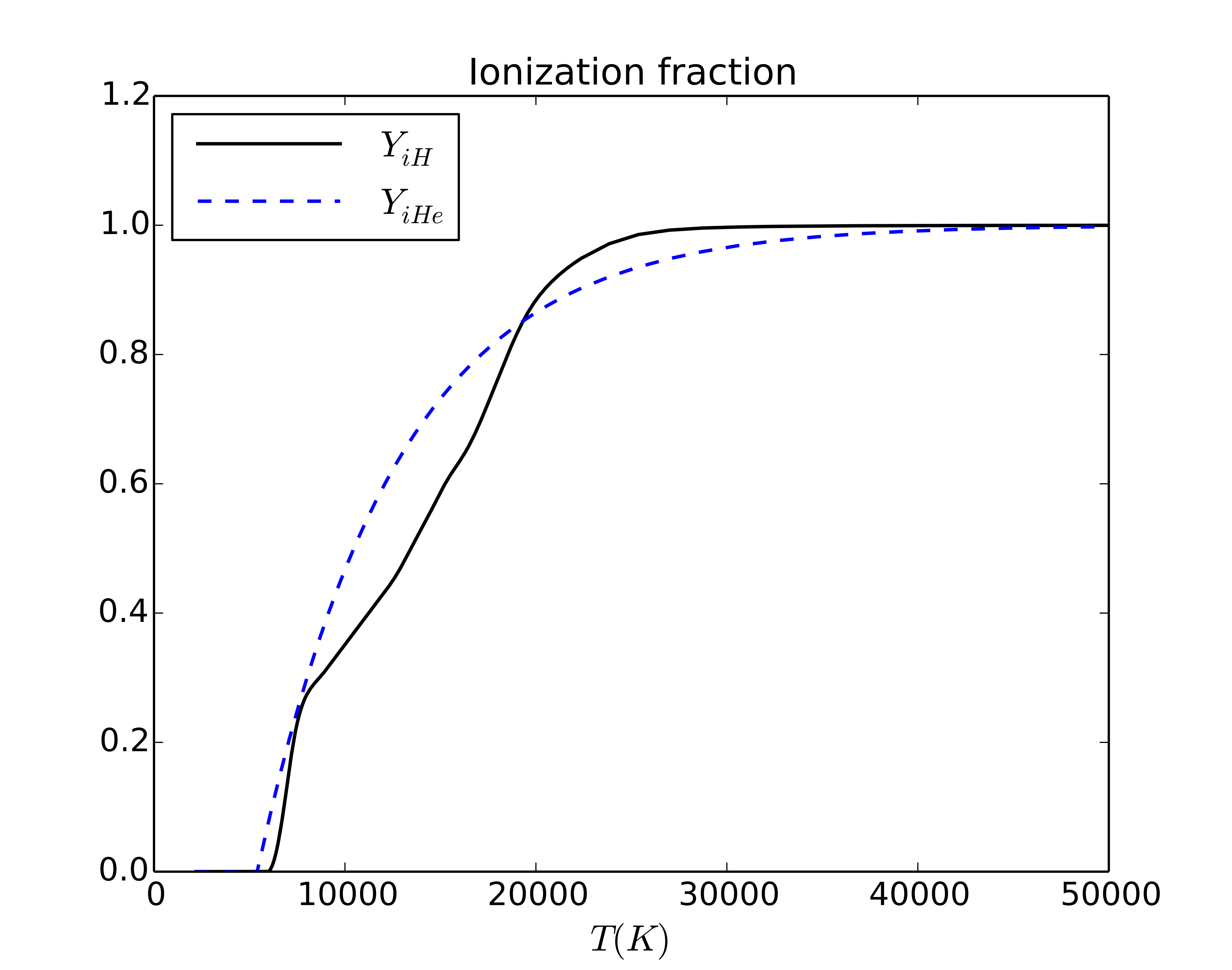}
\caption{Temperature-dependent ionization fractions of hydrogen ($Y_{iH}$) and helium ($Y_{iHe}$).
}\label{f1}
\end{figure}
 
The dynamic viscosity in energy Eq. (\ref{energy}) is given by:
 \begin{equation}
    \xi = \xi_{i}+\xi_{n}=\frac{n_nk_BT}{\nu_{nn}}+\frac{n_ik_BT}{\nu_{ii}},   \label{csi}
 \end{equation}
 where $\xi_i$ and $\xi_n$ are the viscosity coefficients contributed by ions and neutrals, respectively. The collision frequencies $\nu_{nn}$ and $\nu_{ii}$ are given as below \citep{Leake2013}:
\begin{equation}
    \nu_{nn} = n_n\sigma_{nn} \sqrt{\frac{16k_BT}{\pi m_n}},  \label{nunn}
\end{equation} 
\begin{equation}
    \nu_{ii} =  \frac{n_i e_c^4 \Lambda}{3m_i^2\epsilon_0^2} \left( \frac{m_i}{2\pi k_B T} \right)^{3/2}.  \label{nuii}
\end{equation} 
In this work, we simply ignore the contribution from helium for calculating the viscosity coefficients and put the constant values of $e_c$, $m_i$, $\epsilon_0$ and $k_B$ into Eq. (\ref{nuii}), then Eq. (\ref{csi}) can be written as :
 \begin{equation}
   \xi = \xi_{i}+\xi_{n} \simeq \frac{1.63\times10^{-16}}{\Lambda}T^{5/2} +\frac{\sqrt{\pi m_n k_B T}}{4\sigma_{nH-nH}}.
 \end{equation}
The momentum transfer cross section is then used to calculate the viscosity and we take $\sigma_{nH-nH}=1\times10^{-18}$\,m$^2$ and $\Lambda \simeq 10$, then the above equation is calculated as:
 \begin{equation}
    \xi = \xi_{i}+\xi_{n} \simeq 0.671\times10^{-7}\sqrt{T}+1.631\times10^{-17}T^2\sqrt{T}. \label{csif}
 \end{equation} 
Since the viscosity coefficient increases with the plasma temperature and the maximum temperature in the simulation domain in this work is $\sim 8\times10^4$\,K, we can find that the maximum value of the second term in  (\ref{csif}) is normally comparably or much smaller than the first term $0.671\times10^{-7}\sqrt{T}$. Therefore, we simply take:
\begin{equation}
   \xi = 10^{-7}\sqrt{T}.
\end{equation}
 
  \subsection {Radiative cooling models} \label{sec:radiation}
The radiative cooling is very complicated and important in the chromosphere. In order to investigate the effect of radiative cooling on magnetic reconnection, we included two different radiative cooling models ($Q_{rad1}$ and $Q_{rad2}$) in different cases. The chromospheric radiative energy balance is dominated by a small number of strong lines from neutral hydrogen, singly ionized calcium, and singly ionized magnesium \citep{Vernazza1981}. \cite{Carlsson2012} derived a reasonable simple radiative cooling model for the chromosphere based on the three lines. Such a radiative cooling model is applied in this work and given by:

 \begin{equation}
 Q_{rad1} = -\sum_{X=H, Mg, Ca}L_{Xm}(T) E_{Xm}(\tau) \frac{N_{Xm}}{N_X}(T)A_X\frac{N_H}{\rho}n_e\rho,
 \end{equation}
where $L_{Xm}(T)$ is the optically thin radiative loss function varying with temperature $T$, per electron and per particle of element $X$ in ionization stage m, $E_{Xm}(\tau)$ is the escape probability as a function of the depth parameter $\tau$, $\frac{N_{Xm}}{N_X}(T)$ is the fraction of element $X$ that is in ionization stage $m$, $A_X$ is the abundance of element $X$, and $\frac{N_H}{\rho}=4.407\times10^{23}$\,g$^{-1}$ is the number of hydrogen particles per gram of chromospheric material.

Since we assume the current sheet is located in the low solar chromosphere around the solar TMR and parallel to the solar surface in this work, the large optical depth there makes the escape probability of photons from HI line be zero. Therefore, the radiative cooling at such an altitude is mainly contributed by Ca II and Mg II lines. The escape probability of photons from Ca II and Mg II lines are given by $E_{CaII}=1.59\times10^{-2}$ and $E_{MgII}=5.9867\times10^{-2}$ for the region near TMR \citep{Carlsson2012}. We can get the abundances of Ca and Mg as $ A_{Ca}=2.042\times10^{-6}$ and $A_{Mg}=3.388\times10^{-5}$ according to the solar atmosphere model \citep{Avrett2008}. The units in \cite{Carlsson2012} are CGS and we transform the units into that of SI for the purposes of this work. We  only turned on this radiative cooling model when the temperature was higher than $4434$\,K. Otherwise, $Q_{rad1}=0$. We set the heating term $H=0$ when $Q_{rad1}$ is applied to the simulation. Since the initial temperature is lower than $4434$\,K, both the cooling and the heating vanish at $t=0$ in this scenario. 
   
Another radiative cooling model that has also been used in the previous works \citep{Ni2015,Ni2016} is given by \citep{Gan1990}:
  \begin{equation}
  Q_{rad2}= -1.547\times10^{-42}n_en_H \alpha T^{1.5},  
  \end{equation}
 where $n_H$ is the number density of the total hydrogen, $\alpha$ is a parameter that depends on the altitude of the solar atmosphere, equal to about $1.805\times10^{-4}$ at TMR. Initially, we set $H=1.547\times10^{-42}n_{e0}n_{H0} \alpha T_0^{1.5}$ when the $Q_{rad2}$ is applied in the simulation, where $T_0$ is the initial temperature, and $n_{e0}$ and $n_{H0}$ are the initial electron and total hydrogen number densities, respectively. When $t>0$, $H$ is turned off. Such a heating term is included to make $Q_{rad}+H=0$ and the system in equilibrium at the beginning to avoid possible artificial effects in the simulation. Sine the radiative cooling increases with the density of electron and temperature, the initial radiative cooling effect is small compared to that in the late stage of reconnection.

\subsection{Initial setups} \label{sec:setup}
 
In this work, we performed simulations for several cases to investigate the effects of different physical processes and to make sure we have enough resolution in our simulations. The same setup in all the cases are described as follows. The simulation domain extends from 0 to $L_0$ in the $x$-direction and from $-0.5L_0$ to $0.5L_0$ in the $y$-direction, $L_0=2\times10^5$\,m. The adaptive mesh refinement (AMR) skill is applied to simulations, which start with a base-level grid of $192\times192$. The initial temperature is $T_0=4400$\,K, and the initial total mass density is $\rho_0 = 1.66057\times10^{-6}$\,kg\,m$^{-3}$, which are the typical temperature and plasma density around TMR. The horizontal force-free Harris current sheet is used as the initial magnetic configuration in equilibrium (see also, e.g., 
\cite{Ni2015, Ni2016} in all the cases):
\begin{eqnarray}
  B_{x0}&=&-b_0\tanh[y/(0.05L_0)]                               \\
  B_{y0}&=&0   \\
  B_{z0}&=&b_0/\cosh[y/(0.05L_0)].
\end{eqnarray}
The small perturbations of magnetic fields were initialized as below:
\begin{eqnarray}
  b_{x1}&=&-b_{pert} \sin \left[ \frac{2\pi(y+0.5L_0)}{L_0} \right] \cos \left[ \frac{2\pi(x+0.5L_0)}{L_0} \right]                             \\
  b_{y1}&=& b_{pert} \cos \left[ \frac{2\pi(y+0.5L_0)}{L_0} \right] \sin \left[ \frac{2\pi(x+0.5L_0)}{L_0} \right] ,
\end{eqnarray}
where $b_{pert}=0.05b_0$. The initial velocities set to zero. The open boundary conditions are used in the simulations.

In the next section, only the results in four cases are presented. The differences among the four cases are listed as follows. The radiative cooling model $Q_{rad1}$ is applied in Cases I, III, and IV, with the radiative cooling model $Q_{rad2}$ applied in Case II.  The only difference between Case I and Case III is the initial strength of magnetic fields, $b_0$, such that $b_0=0.05$\,T in cases I and II, $b_0=0.09$ in cases III and IV. The highest level of adaptive mesh refinement in case I, II, and III is 9, and the corresponding smallest grid size is $2$\,m. Such a high resolution allows the numerical diffusion to be smaller than the physical diffusions in case I, II, and III. The low resolution in Case IV causes the numerical diffusion to dominate in the magnetic reconnection process. Details are presented in Table.1.  
 
\begin{table*}[]
\centering
\caption{Differences among four simulation cases.}
\label{tab1}
\begin{tabular}{lclcccc}
\hline
\multirow{2}{*}{Caes} & \multirow{2}{*}{\begin{tabular}[c]{@{}c@{}}Highest\\ AMR Levels\end{tabular}} & \multirow{2}{*}{\begin{tabular}[c]{@{}c@{}}Radiative\\ Cooling\end{tabular}} & \multirow{2}{*}{\begin{tabular}[c]{@{}c@{}}Ambipolar \\ Diffusion\end{tabular}} & \multirow{2}{*}{\begin{tabular}[c]{@{}c@{}} Initial magnetic field \\$b_0$(\,T) \end{tabular}}  \\
     & & & & \\ \hline
I    & 9 & $Q_{rad1}$ & Yes & 0.05  \\ \hline
II   & 9 & $Q_{rad2}$ & Yes  & 0.05  \\ \hline
III  & 9 & $Q_{rad1}$ & Yes & 0.09  \\ \hline
IV  & 0 & $Q_{rad1}$ & Yes & 0.09  \\ \hline
%V   & 9 & $Q_{rad2}$ & Yes & 0.05  \\ \hline
%VI  & 9 &       No         & Yes & 0.05  \\ \hline
\end{tabular}
\end{table*}

%--------------------------------------------------------------------------------------------------------------------------------%
\section{Results and discussions} \label{sec:results}
%-----------------------------------------------------------------
\subsection{Effects of radiative cooling} \label{sec:resolution}

Whether the low solar chromosphere can be heated to a temperature up to $20,000$\,K or even higher is still an open question. This reflects an important issue related to the formation height of the UV burst. In this work, we study magnetic reconnection in a low $\beta$ environment around TMR with more realistic diffusions and radiative cooling.

Figure 2 shows distributions of temperature and magnetic field in the $x$-$y$ plane for both Case I and Case II. Looking into the evolutions in the magnetic field, we find many magnetic islands forming in the unstable reconnection process during the later stage in both cases. These islands have a closed loop with an O-point in the center and they coalesce with the nearby one to grow bigger and bigger. They are also known as plasmoids. Comparing the left and the right panel in Figure 2, we can find that the temperature distributions are almost the same in the two cases before the unstable reconnection process marked by the formation of plasmoids. 

After plasmoid instability takes place, more plasma is heated to high temperatures of about $20,000$\,K in Case II, with the simple radiative cooling model $Q_{rad2}$. Only a small fraction of plasma in Case I is heated to  high temperatures about $20,000$\,K till $t=10.76$\,s. However, the maximum temperature reaches a higher value in Case I during the later stage; the maximum temperature reaches above $50,000$\,K in Case I and it is only about $30,000$\,K in Case II at around $t=10.5$\,s, as shown in Figure 5(a). Cases I and II are both terminated before $t=11$\,s because of the extremely small time step, it is possible that more plasma could be heated to higher temperature if the simulation could last longer. As shown in Figure 2, the high temperature plasma is usually located in the regions surrounding plasmoids. 

Figure 3 presents the distributions of plasma density in case I and II, which shows that the plasma density becomes nonuniform in the reconnection region after plasmoid instability turns on, and this phenomenon is more obvious in Case I. The dense plasma is concentrated in the head regions of the plasmoids in both cases. The maximum hydrogen density exceeds $4\times10^{21}$\,m$^{-3}$ in Case I at $t=10.76$\,s, but the maximum hydrogen density is much lower in Case II at about the same time. Comparing Figures 2 and 3, we can find that the high temperature above $20,000$\,K appears in the regions with the hydrogen density in the range of $\sim(2-5)\times10^{20}$ m$^{-3}$. 

Figure 4 displays the distributions of the plasma density, temperature, and pressure along the vertical white dashed-dotted line shown on the bottom panel in Figures 2(a) and 3(a). This line crosses a pair of shocks surrounding the big plasmoid, the temperature reaches the peak values around the two shock fronts, the plasma density and pressure sharply increase behind the two shock fronts. Therefore, the temperature increases in these  regions are likely due to the shock compression.

Figure 5 shows the time dependent maximum temperature, $T_{max}$, the maximum velocity in the $x$-direction, $v_{x-max}$, the maximum velocity in the $y$-direction, $v_{y-max}$, and the maximum current density in the $z$-direction $J_{z-max}$. We can see that the evolution in these variables in case I is very similar to that in Case II, even after the plasmoid instability appears. However $T_{max}$ in Case I starts to become larger than Case II after $t=9.5$\,s.

Comparing the results in cases I and II, we conclude that the radiative cooling affects the distributions of the plasma temperature and density in an apparent fashion after plasmoid instability takes place. The radiative cooling model $Q_{rad1}$ from \cite{Carlsson2012} generates fewer plasmas to be heated to about $20,000$\,K when the unstable magnetic reconnection process evolves to the same time. The difference between the maximum and minimum densities in Case I with $Q_{rad1}$ is larger than that in Case II with $Q_{rad2}$, so does the difference between the maximum and minimum temperatures. In both cases, heating the plasma in the low chromosphere to the temperature up to $2\times 10^{4}$~K or even above is always possible by turbulent reconnection in the low $\beta$ environment. 

As in case II, magnetic reconnection of the same plasma $\beta$ and the radiative cooling model was also studied by Ni et al. (2016), in which the plasma could be heated up to $10^{5}$~K as the ionization fraction is a constant. In the present work, the temperature-dependent ionization fractions of hydrogen and helium enhance the radiative cooling in the reconnection region by several orders of magnitude, which suppresses the maximum temperature to an apparently low value, say $3\times 10^{4}$~K. This means that including the temperature-dependent ionization fraction in the simulation is a measure that allows us to duplicate more realistic evolutionary behaviors of the temperature in the reconnection process. 
 
\begin{figure*}
    \centerline{\includegraphics[width=0.45\textwidth, clip=]{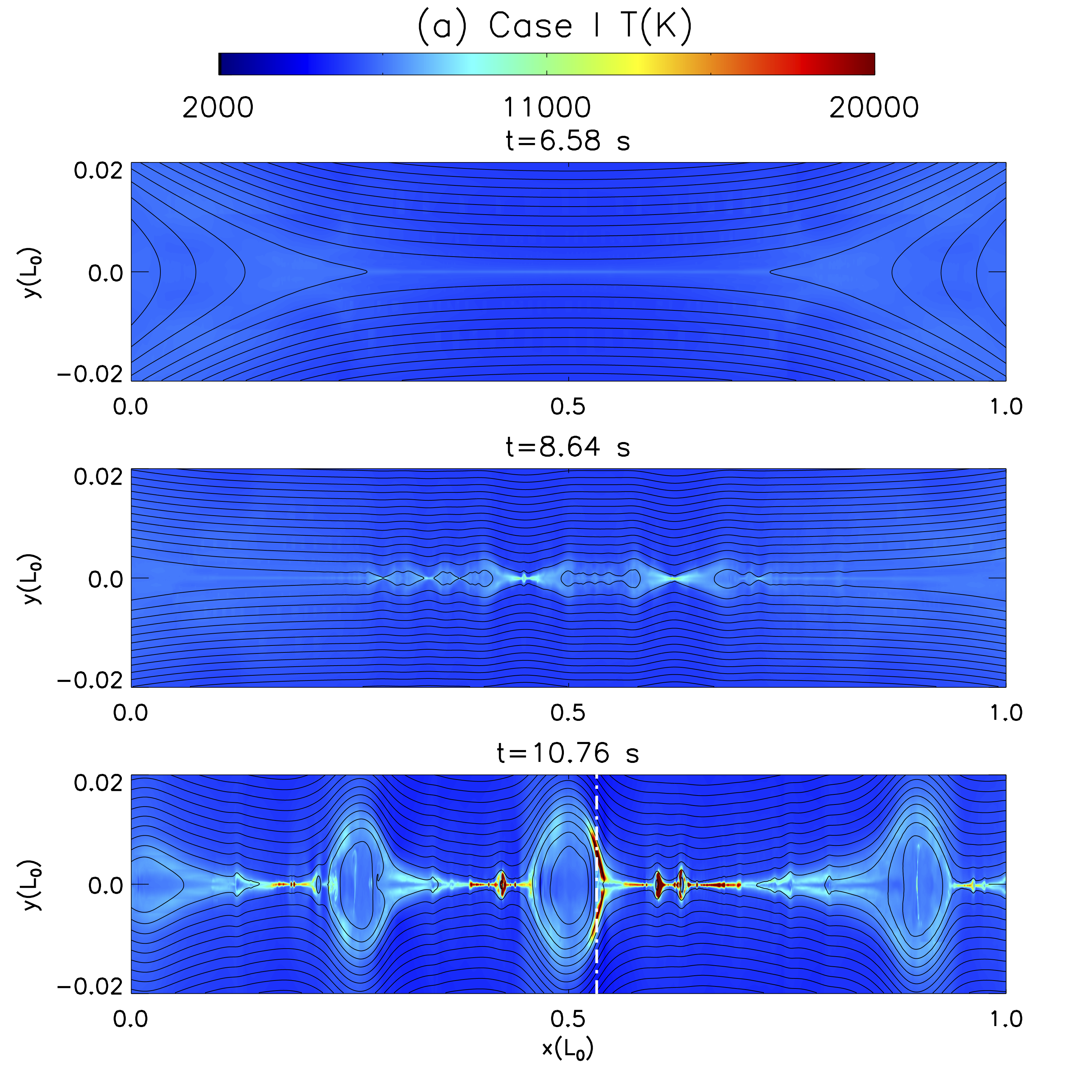}
                       \includegraphics[width=0.45\textwidth, clip=]{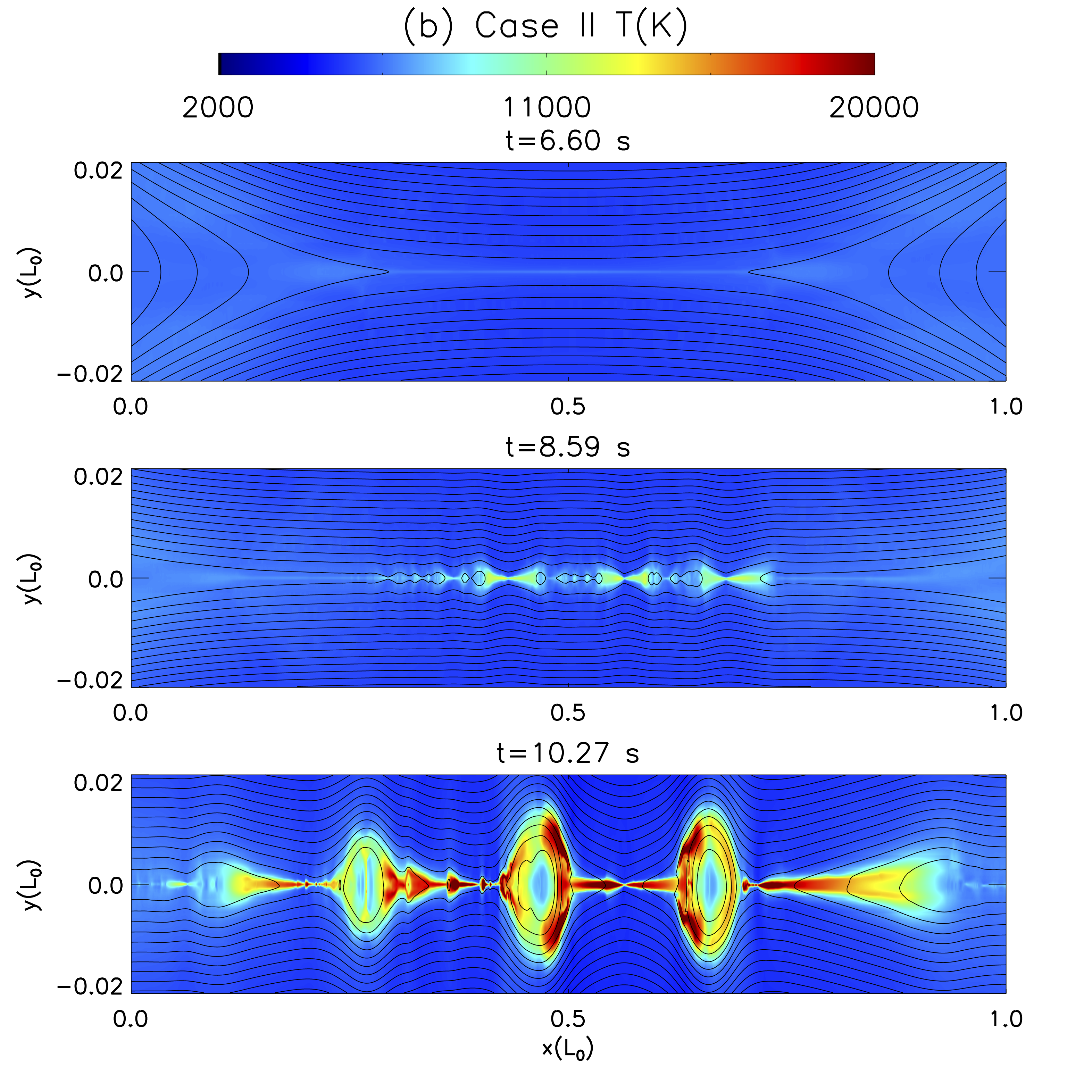}} 
  \caption{Distributions of temperature (background color) and magnetic field lines (black solid lines) at three different times in Case I (a) and Case II (b). } 
  \label{f2}
 \end{figure*}

\begin{figure*}
    \centerline{\includegraphics[width=0.45\textwidth, clip=]{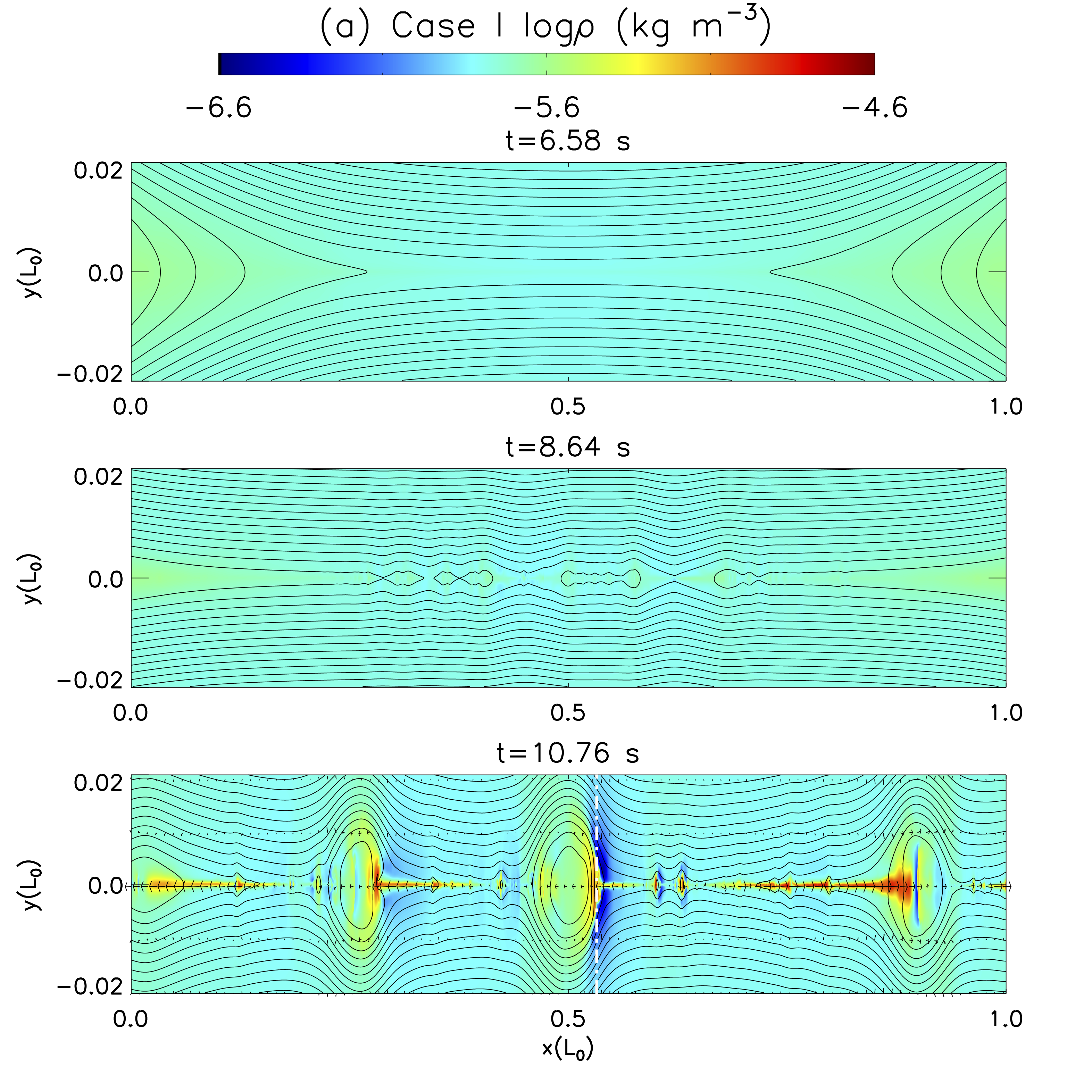}
                       \includegraphics[width=0.45\textwidth, clip=]{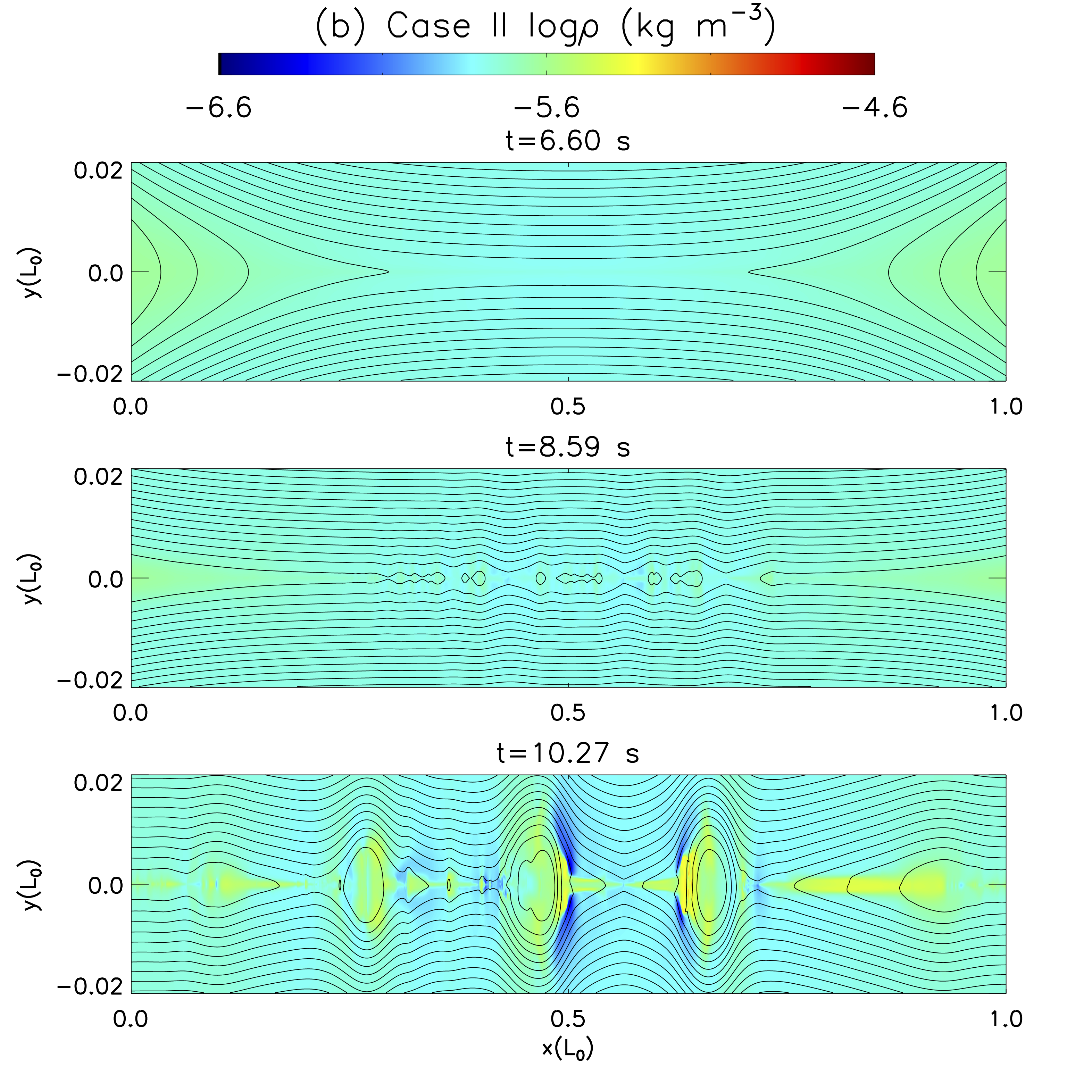}} 
  \caption{ Distributions of mass density (background color) and magnetic field lines (black solid lines) at three different times in Case I (a) and Case II (b). The black arrows on the bottom panel in (a) represent the velocity.}
   \label{f3}
\end{figure*}

\begin{figure}
    \centerline{\includegraphics[width=0.45\textwidth, clip=]{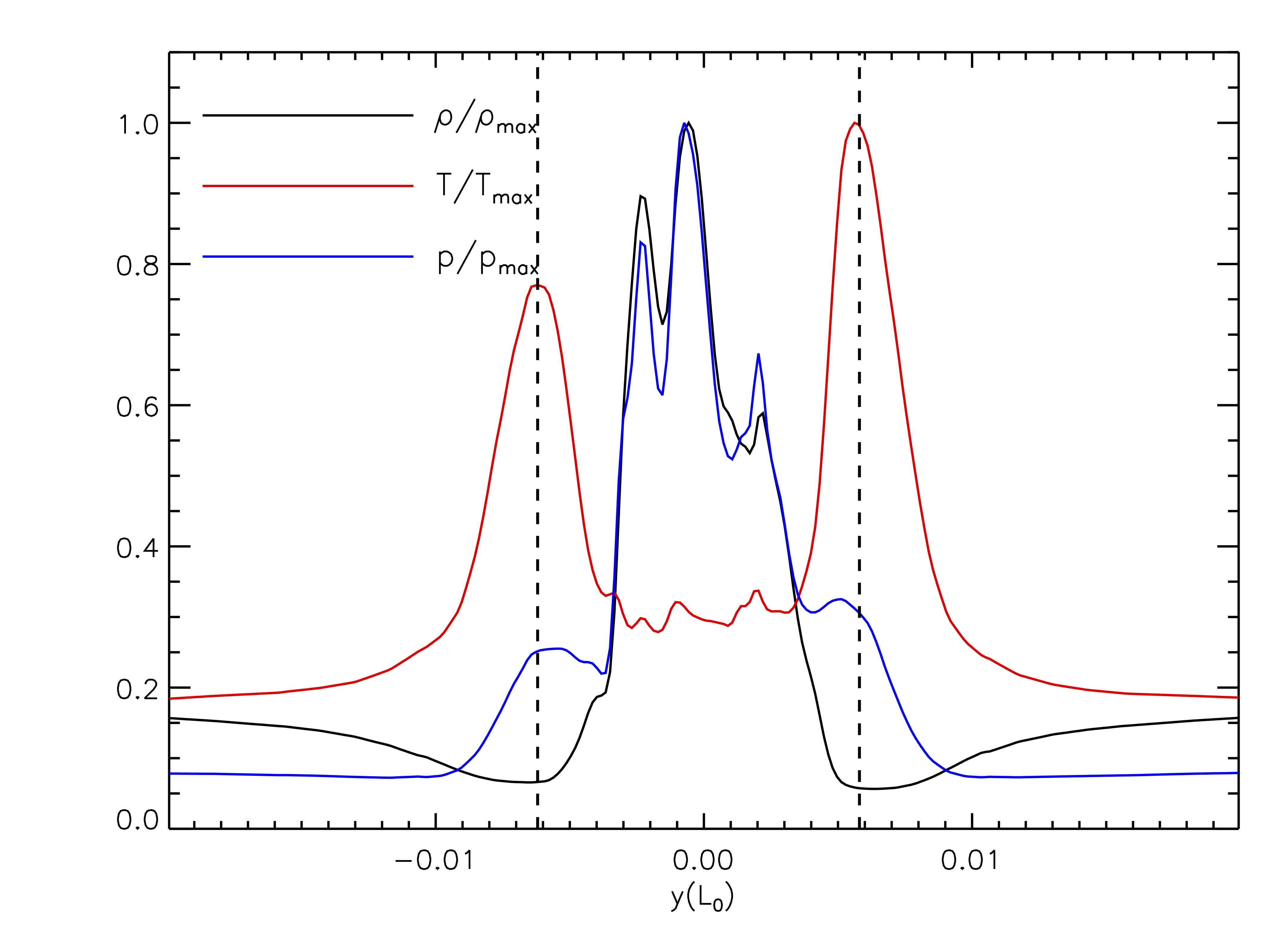}} 
  \caption{Distributions of the mass density ($\rho/\rho_{max}$), temperature ($T/T_{max}$), and pressure ($p/p_{max}$) along the white dashed-dotted line as shown on the bottom panel of Figure. 2(a) and Figure. 3(a); $\rho_{max}=8.36\times10^{-6}$\,kg\,m$^{-3}$, $T_{max}=24504$\,K, and $p_{max}=494.55$\,N\,m$^{-2}$ are the maximum values of mass density, temperature, and pressure along this line. }
   \label{f4}
\end{figure}

\begin{figure*}
    \centerline{\includegraphics[width=0.35\textwidth, clip=]{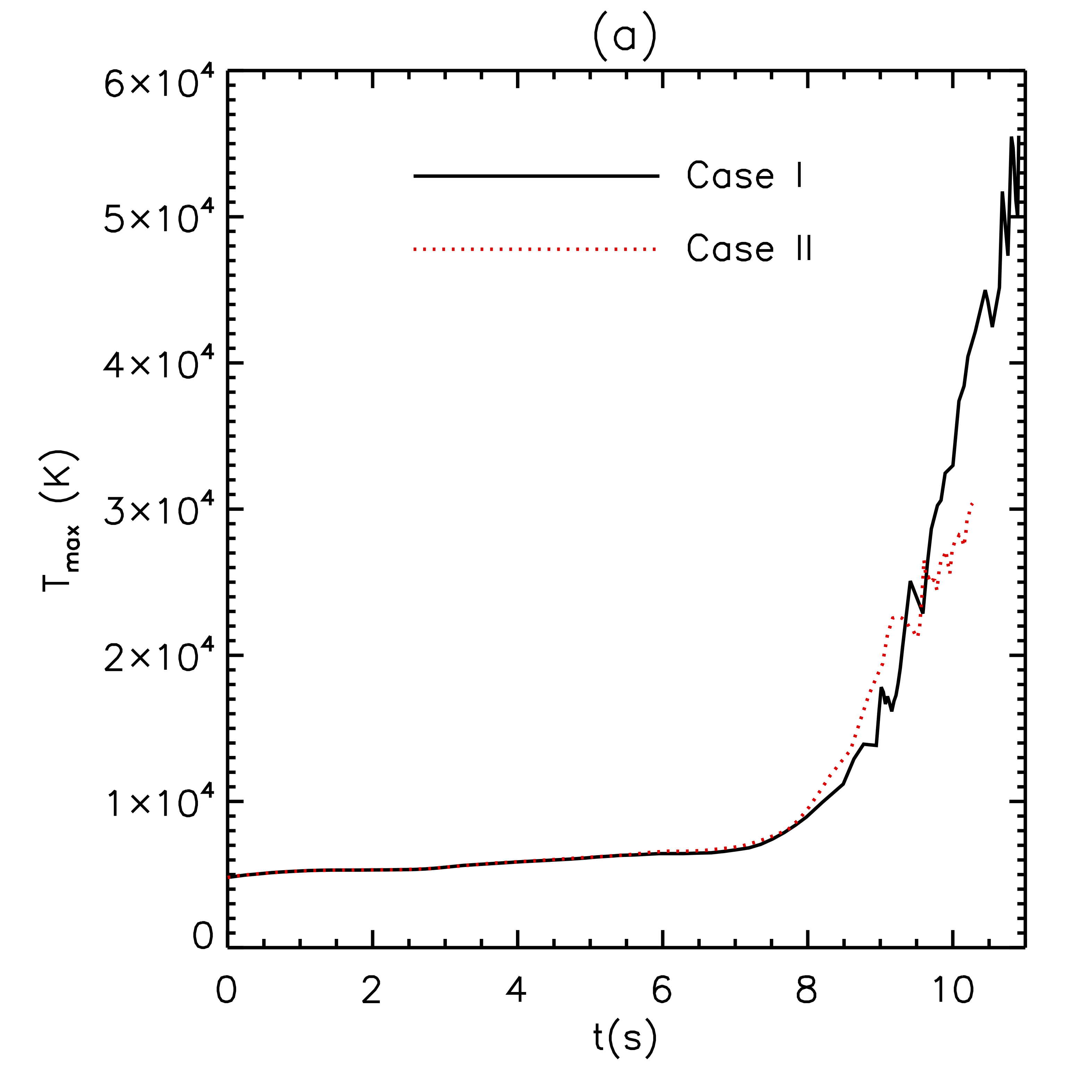}
                      \includegraphics[width=0.35\textwidth, clip=]{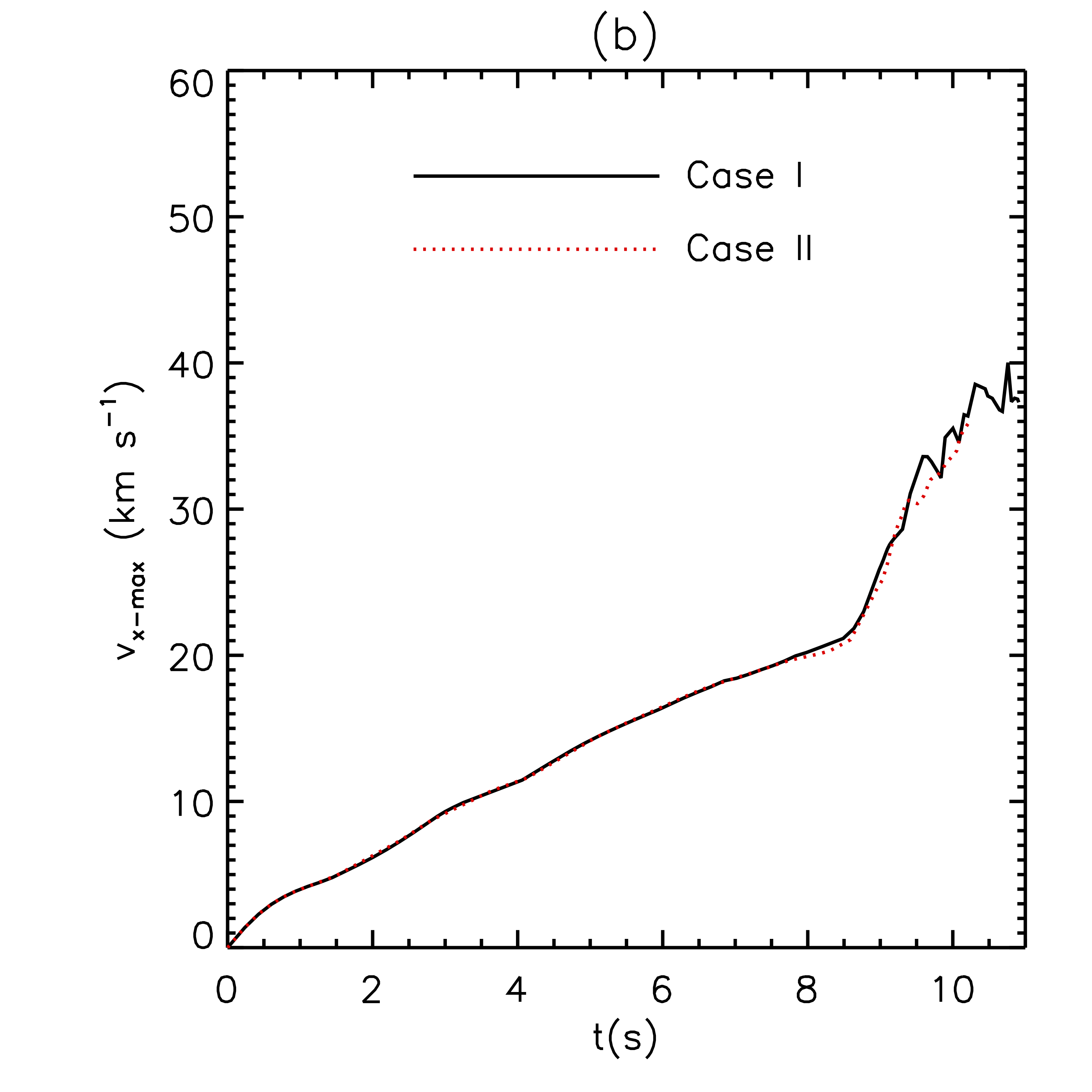}} 
    \centerline{\includegraphics[width=0.35\textwidth, clip=]{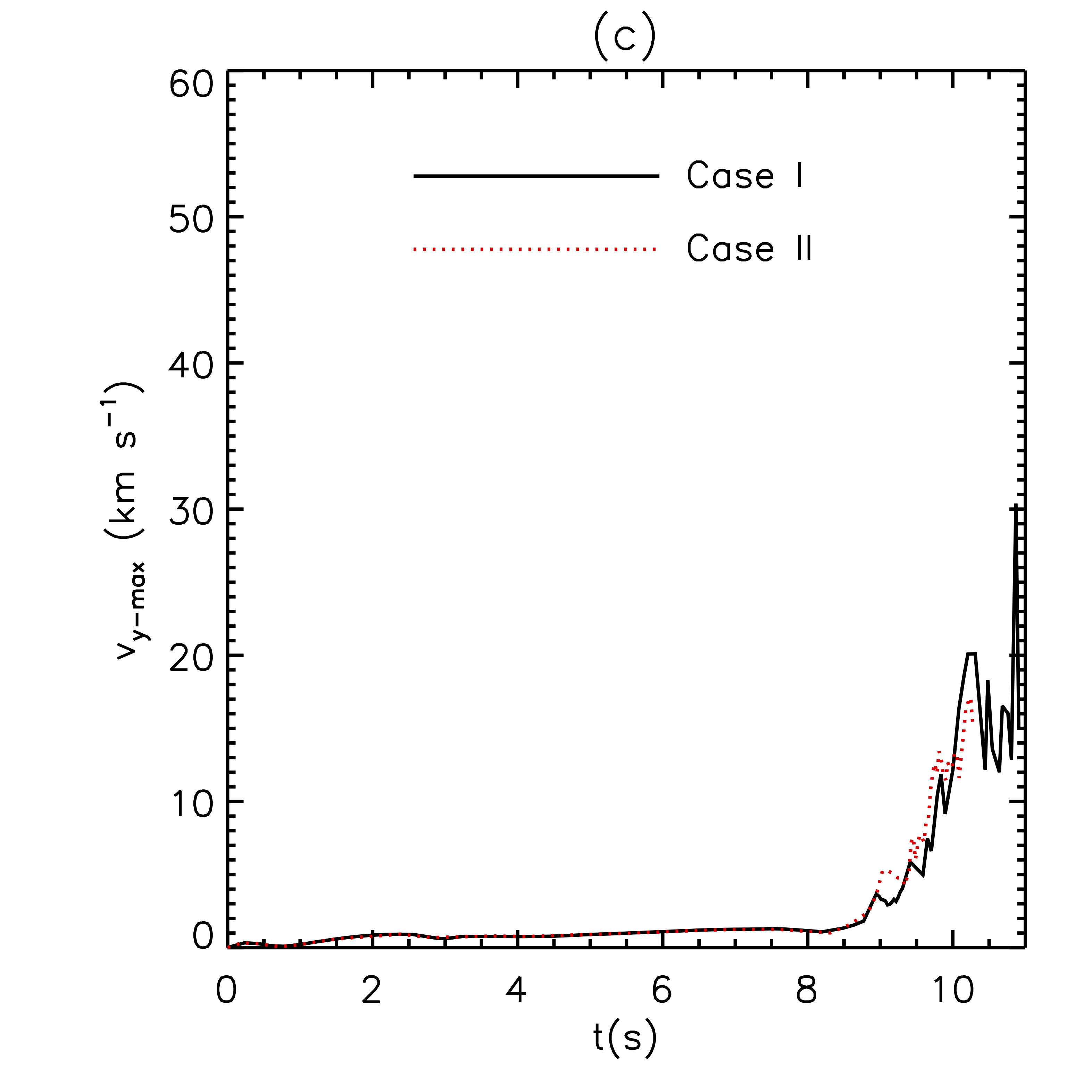}
                       \includegraphics[width=0.35\textwidth, clip=]{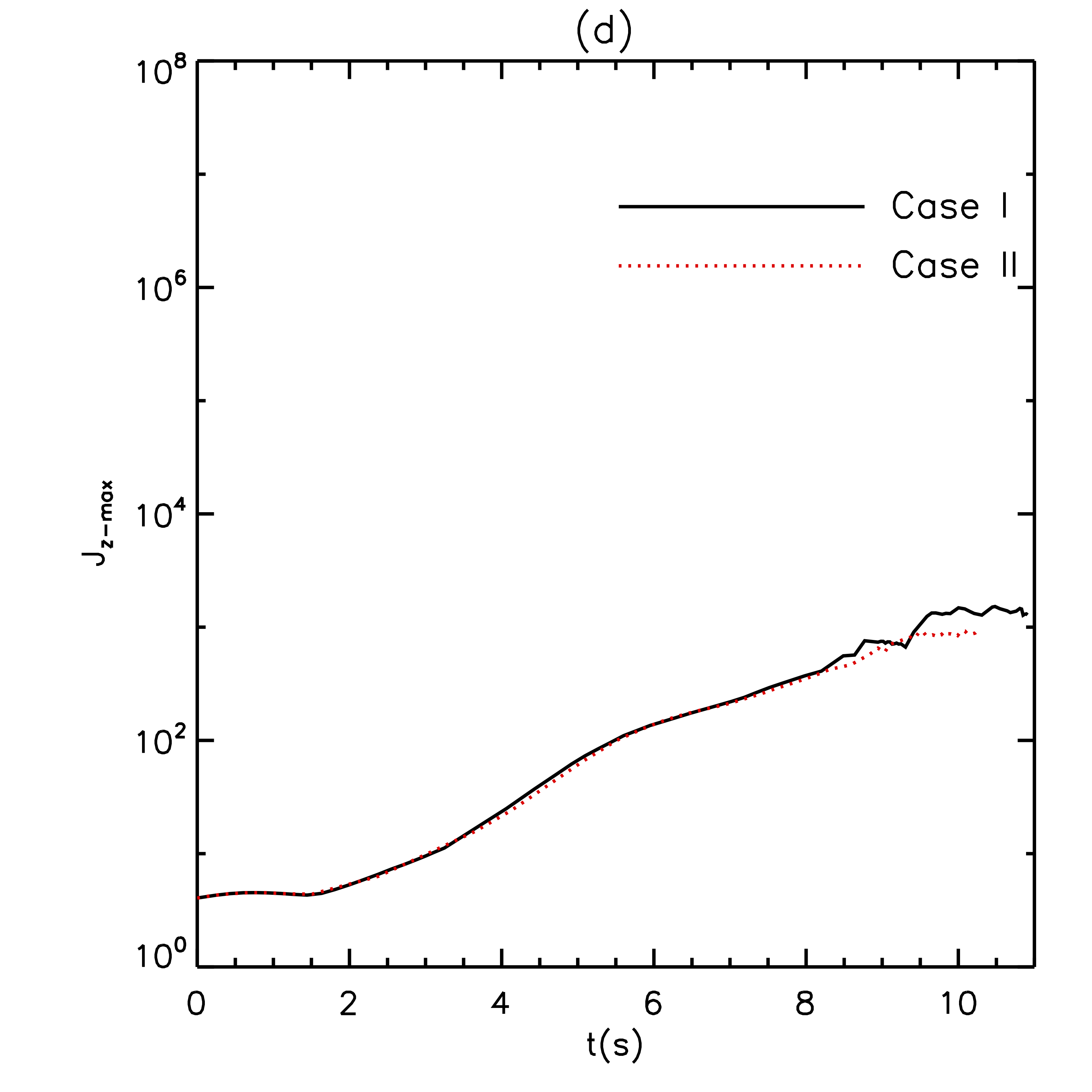}}                    
 \caption{Evolutions of the maximum temperature $T_{max}$ (a), the maximum velocity in the $x$-direction $v_{x-max}$ (b), the maximum velocity in the $y$-direction $v_{y-max}$ (c), and the maximum current density in the $z$-direction $J_{z-max}$ (d) with time in Case I and Case II. }
   \label{f5}
\end{figure*}

\subsection{Mechanisms for conversion of magnetic energy into heating} \label{sec:terms}

In this portion of the work, we present and discuss the mechanisms for converting magnetic energy into heating via reconnection in the low $\beta$ environment near TMR. Figure 6 displays the time-dependent diffusivity at the principal X-point (PX-point) inside the current sheet in cases I and II. In order to unify the units of three diffusion coefficients, we evaluate the ambipolar diffusion in the way of $\eta_{amp} = \eta_{AD}B^{2}/\mu_{0}$. Initially, the ambipolar diffusion $\eta_{Amp}$ is more than one order of magnitude greater than the magnetic diffusion caused by electron-neutral collision, $\eta_{en}$, while $\eta_{en}$ is more than one order of magnitude greater than the magnetic diffusion caused by electron-ion collision $\eta_{ei}$. These diffusivities at the PX-point all decrease with time, but $\eta_{Amp}$ and $\eta_{en}$ decrease much faster than $\eta_{ei}$. When the plasmoid instability is invoked ($\sim 7$\,s), $\eta_{Amp}$ decreases to a value smaller than $\eta_{ei}$ at the PX-point. Eventually, $\eta_{en}$ and $\eta_{Amp}$ are both more than one order of magnitude smaller than $\eta_{ei}$ at the PX-point. In comparing Figures 6(a) and 6(b), we realize that the evolution in these diffusivities at the PX-point are very similar to one another in cases I and II.

Figure 7 shows the evolution in the average power densities contributed by different heating mechanisms in the reconnection region in cases I and II. Variables $Q_{ei}$, $Q_{en}$, $Q_{Amp}$, $Q_{vis}$, and $Q_{comp}$ in Figures 7(a) and 7(c) are the average power densities of the Joule heating contributed by the magnetic diffusions $\eta_{ei}$ and $\eta_{en}$, the average power densities of the heating contributed by ambipolar diffusion, $\eta_{AD}$, the viscosity, $\xi $, and  compression, respectively. These are evaluated as: $Q_{ei}=\int_{y_s}^{y_e} \int_{x_s}^{x_e}\frac{\eta_{ei}}{\mu_0}|\nabla \times \mathbf{B}|^2dx dy/A_{rec}$, $Q_{en}=\int_{y_s}^{y_e} \int_{x_s}^{x_e}\frac{\eta_{en}}{\mu_0}|\nabla \times \mathbf{B}|^2 dx dy/A_{rec}$, $Q_{Amp}=\int_{y_s}^{y_e} \int_{x_s}^{x_e}\frac{\eta_{AD}}{\mu_0^2}|\mathbf{B} \times (\nabla \times \mathbf{B})|^2 dx dy/A_{rec}$, $Q_{vis}=\int_{y_s}^{y_e} \int_{x_s}^{x_e} \frac{1}{2\xi}Tr(\tau_S^2) dx dy/A_{rec}$ and $Q_{comp}=\int_{y_s}^{y_e} \int_{x_s}^{x_e} (-p\nabla \cdot \mathbf{v})dx dy/A_{rec}$, where $x_s=0.25L_0$, $x_e=0.75L_0$, $y_s=-0.005L_0$, $y_e=0.005L_0$ and $A_{rec}=(x_e-x_s)\cdot (y_e-y_s)=0.005L_0^2$ is the area of the main reconnection region, $Tr$ means evaluating the trace of a matrix. The average power density of the plasma bulk kinetic energy $Q_{kin}$ in Figures 7(b) and 7(d) is calculated as $Q_{kin}=\int_{y_s}^{y_e} \int_{x_s}^{x_e} \frac{\partial (0.5 \rho \mathbf{v}^2)} {\partial t} dx dy/A_{rec}$.

From Figures 7 (a) and 7(c), we notice that $Q_{en}$ is larger than $Q_{ei}$ and $Q_{Amp}$ before $t=3.5$\,s. Therefore, the Joule heating as a result of the electron-neutral collision dominates the process in which the magnetic energy is directly converted into the thermal energy at the very beginning of the magnetic reconnection. As the temperature continuously increases with time inside the reconnection region, $Q_{en}$ decreases and $Q_{ei}$ sharply increases. Eventually, $Q_{ei}$ is much larger than $Q_{en}$ and $Q_{Amp}$. Small values of $Q_{Amp}$ in the whole reconnection process indicate that the ambipolar diffusion is not important for heating in the low $\beta$ reconnection process similar to the case studied previously \citep{Ni2021}, so does the viscous heating occurring here. However, we notice that the compression heating is very important, $Q_{comp}$ sharply increases to a value about one order of magnitude higher than $Q_{ei}$ after plasmoid instability takes place. As shown in Figures 7(b) and 7(d), $Q_{kin}$ subsequently decreases when $Q_{comp}$ continually increases, which indicates that the bulk kinetic energy generated in the reconnection process is subsequently converted into thermal energy by the compression.

Comparing the results in cases I and II, we find that the evolution of all the six components shown in Figure 7 is very similar in the two cases. Therefore, different radiative cooling models do not significantly affect the heating mechanisms in the low $\beta$ reconnection studied here. These results demonstrate that the local compression heating is the dominant mechanism for heating in the reconnection process accounting for the UV burst event in the low chromosphere, the magnetic energy directly converted into thermal energy by magnetic diffusion or ambipolar is much lower. 

Figure 7 also indicates that the total thermal energy in the reconnection process sharply increases with time, especially during the later stage when the plasmoids become bigger and more fragmental current sheets are created. The plasmoid instability  that results in a turbulent reconnection process and many reconnection X-points, enlarges the reconnection region and speeds up the dissipation of magnetic energy and the generation of thermal energy. Therefore, we can expect that the average power density $Q=Q_{comp}+Q_{ei}+Q_{en}+Q_{Amp}+Q_{vis}$ by all the heating mechanisms will increase to a larger value than $2000$\,erg\,cm$^{-3}$\,s$^{-1}$ if the simulation can last for a longer time. 

According to our calculations in Figure 7, we take the average power density $Q=1000$\,erg\,cm$^{-3}$\,s$^{-1}$ as shown in Case I and assume that the size of a typical UV burst is $700 \times700\times700$\,km$^3$ and the life time is $10$\,min, then we can get the total thermal energy generated in the reconnection region of such a UV burst is $2\times10^{29}$\,erg, which is similar to that obtained in a typical UV burst event from observations \citep{Peter2014, Young2018}. Therefore, our results demonstrate that heating in the reconnection process studied in this work could indeed account for the UV burst in the low chromosphere. We note here that this simple calculation was performed on an assumption of the invariant physical condition in the vertical direction to evaluate the total thermal energy released during the UV burst. In reality, if the physical condition, (e.g., the plasma density) changes violently in the vertical direction, the mechanism for reconnection, (and consequently that for heating) might vary from altitude to altitude. This scenario needs to be investigated via 3D simulations in the future.

\begin{figure*}
    \centerline{\includegraphics[width=0.40\textwidth, clip=]{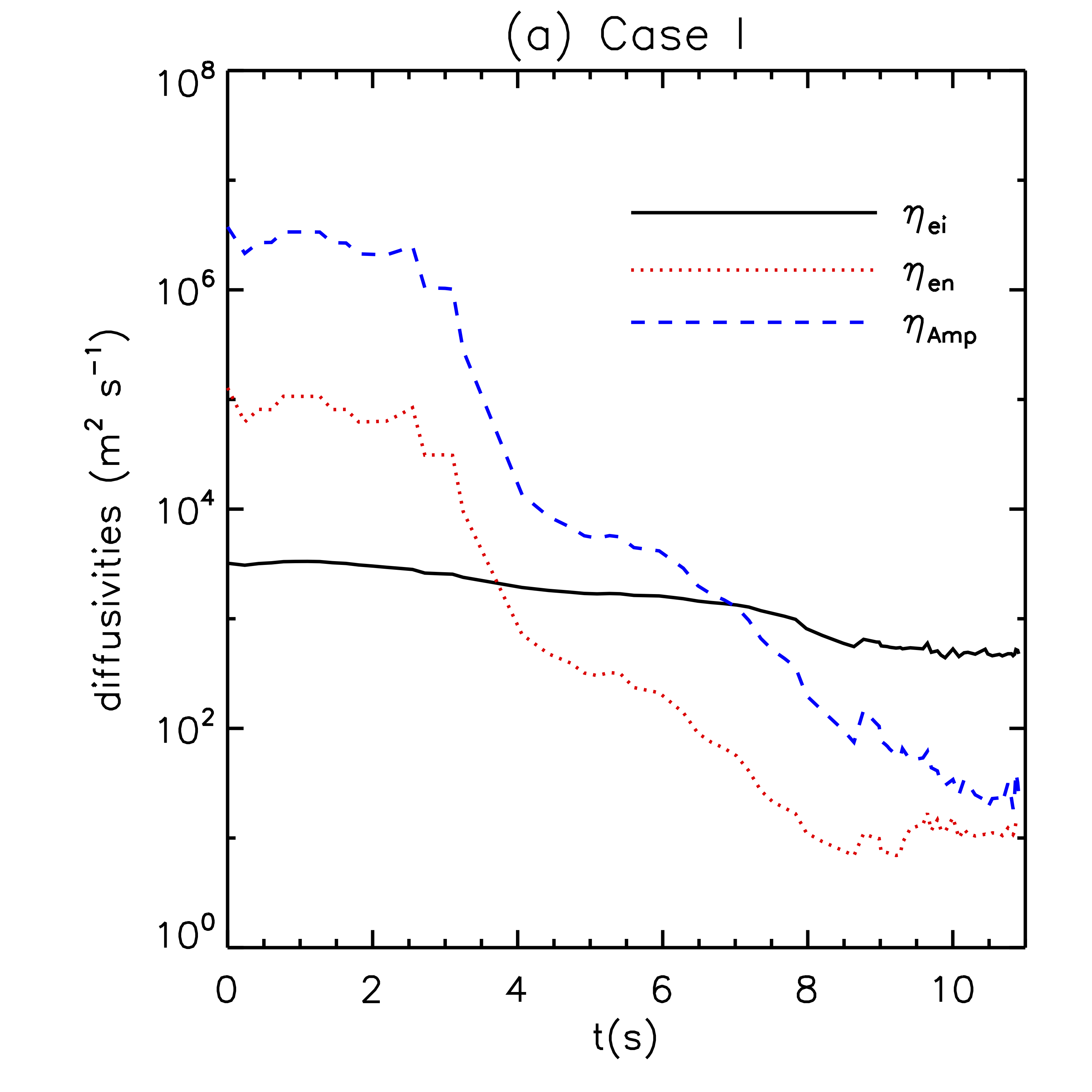}
                       \includegraphics[width=0.40\textwidth, clip=]{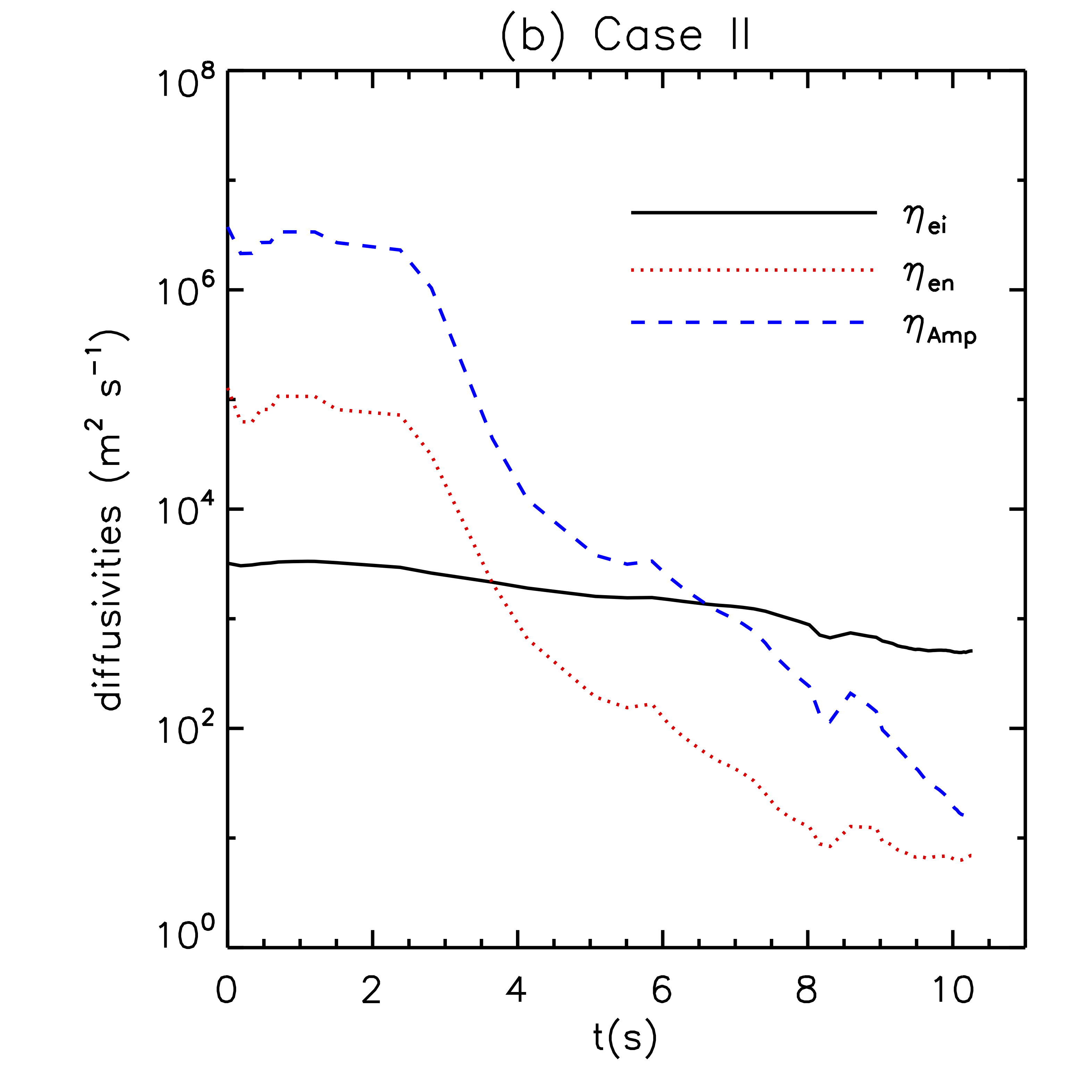}} 
  \caption{Evolutions of different diffusivities at the reconnection PX-point with time in Case I and Case II. } \label{f6}
\end{figure*}

\begin{figure*}
 \centerline{\includegraphics[width=0.35\textwidth, clip=]{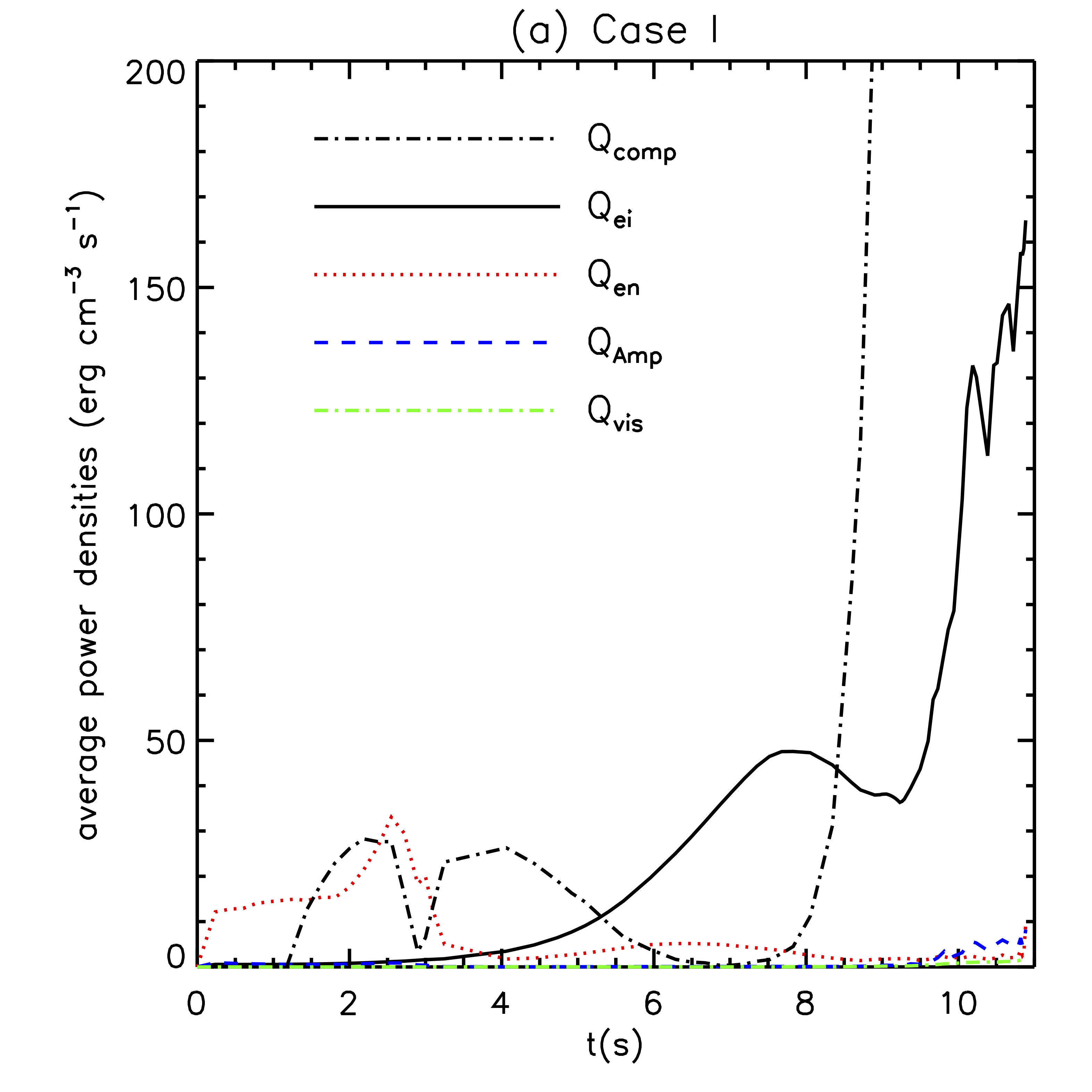}
                      \includegraphics[width=0.35\textwidth, clip=]{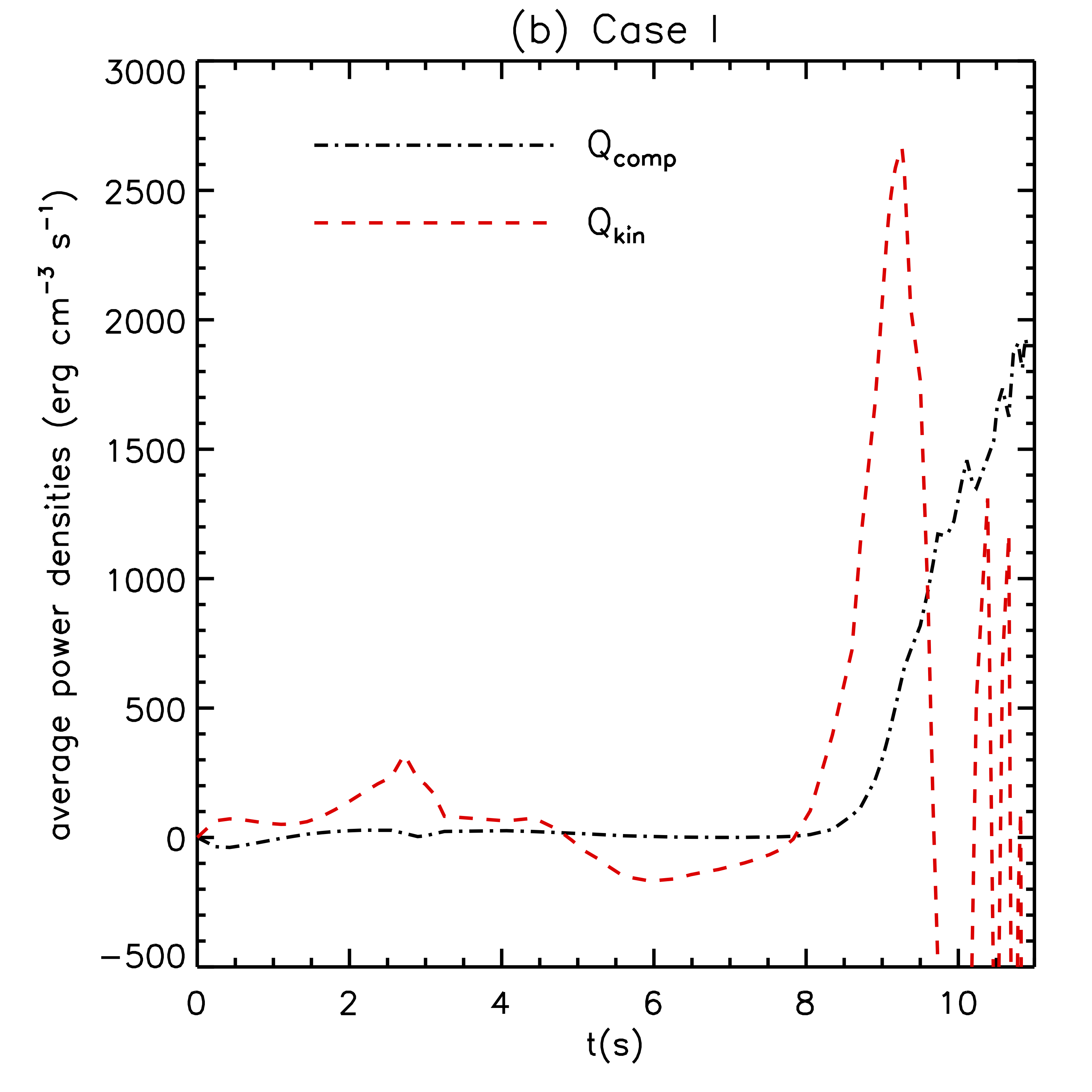}} 
    \centerline{\includegraphics[width=0.35\textwidth, clip=]{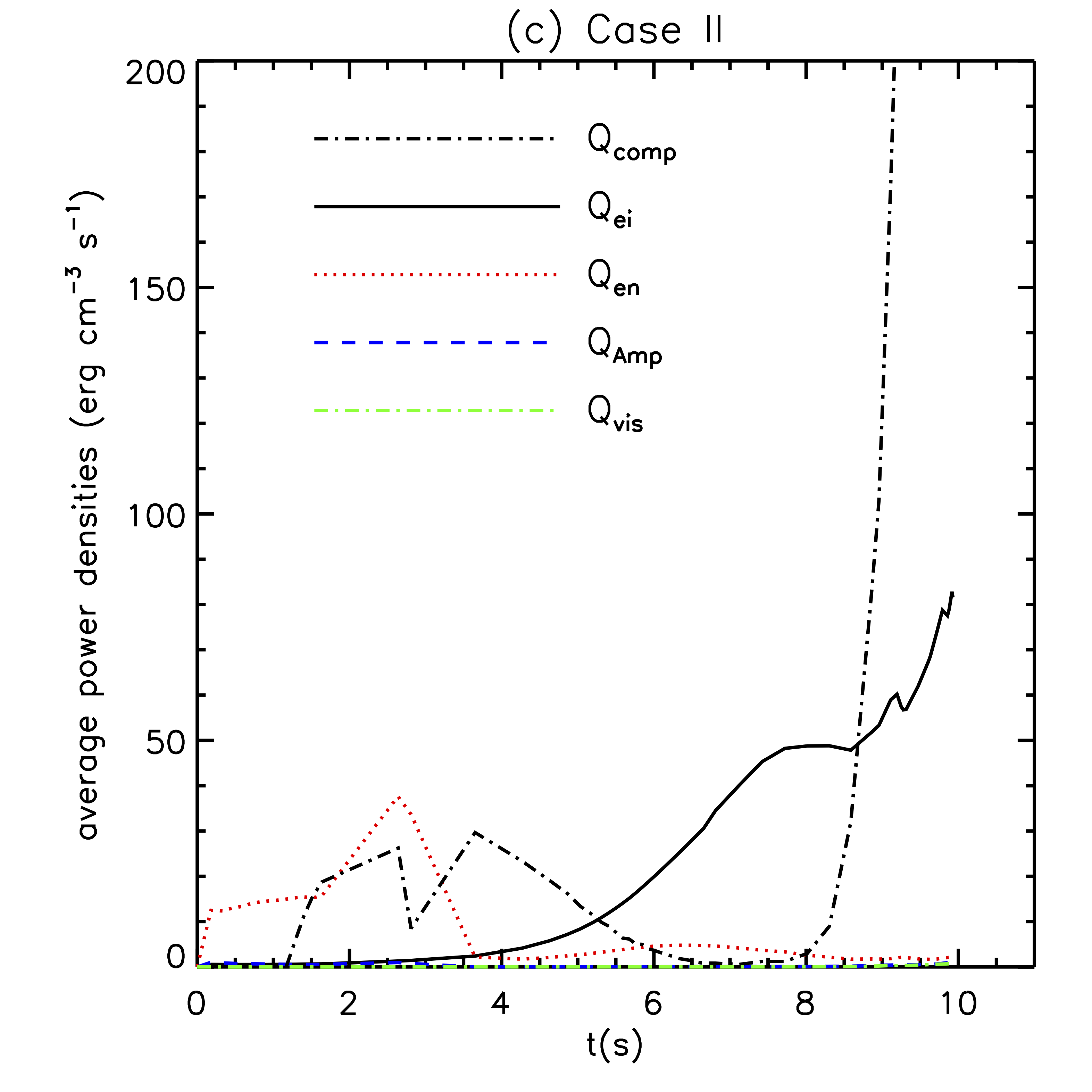}
                       \includegraphics[width=0.35\textwidth, clip=]{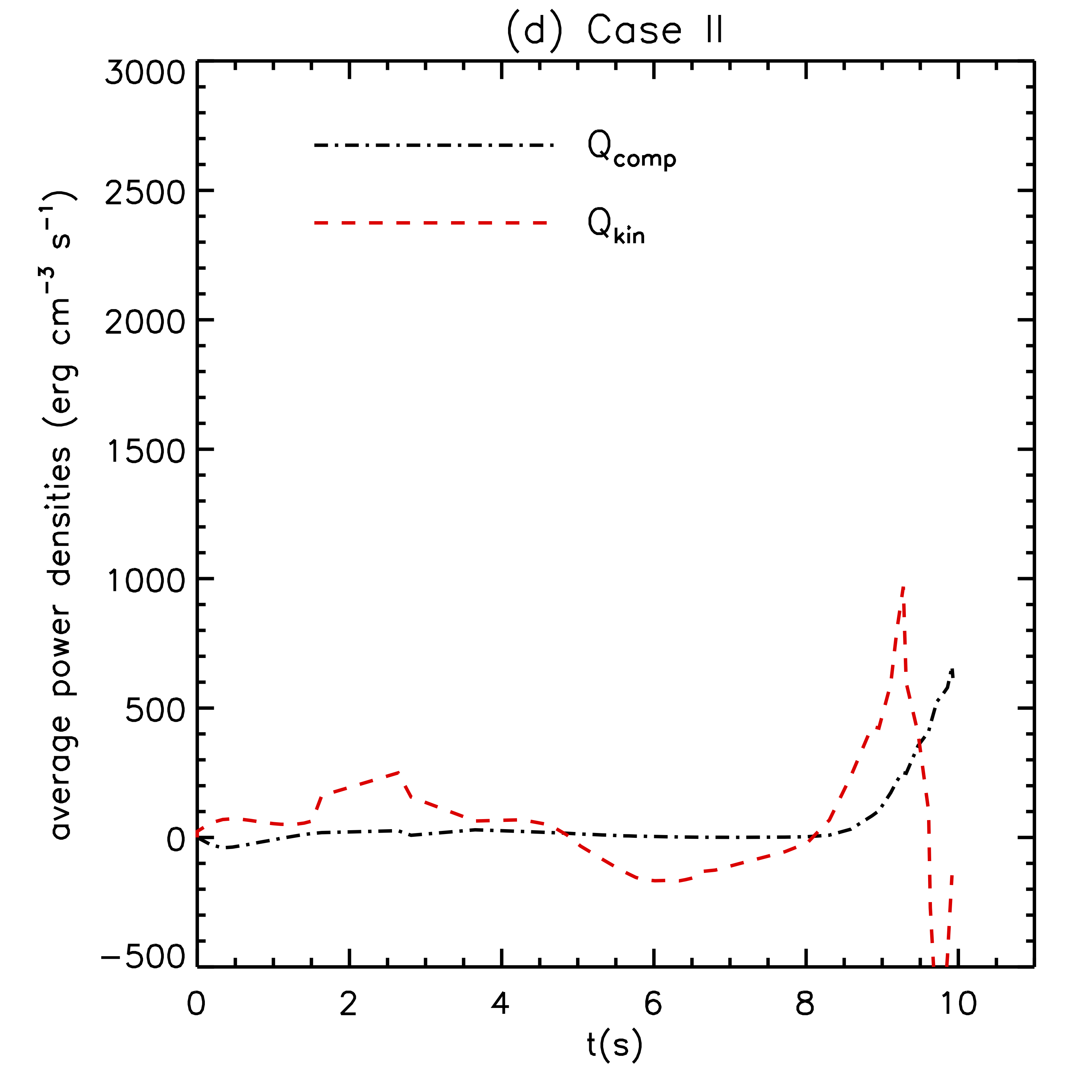}} 
  \caption{Evolutions of different average power densities with time in the reconnection region in Case I and Case II.} 
  \label{f7}
\end{figure*}

%-----------------------------------------------------------------
\subsection{Evolution in synthetic Si~{\sc{iv}} line profiles} \label{sec:spec}

We created the synthesized Si~{\sc{iv}} 1394 {\AA} line profiles for case I, II and III based on the method described in \citet{Peter2006} using the atomic data package CHIANTI \citep[version 9.0;][]{Dere2019}. The high-temperature plasma appearing in case I and II has strong response to Si~{\sc{iv}} emission, but the reconnection outflow velocity in cases I and II is not fast enough to generate broad profile of the Si~{\sc{iv}} spectral line as shown by observations. Therefore, we simulated Case III with a much stronger reconnection magnetic field to get the broader Si~{\sc{iv}} spectral line profile. Two dimensional distributions of the temperature and the velocity in the x-direction at three different times are presented in Figure 8. Figure 8(a) shows that lots of plasma is already heated to higher temperatures above $40,000$\,K at $t=5.74$\,s. The maximum outflow velocity reaches $\sim70$\,km\,s$^{-1}$ as shown in Figure 8(b). 

The synthesized Si~{\sc{iv}} 1394 {\AA} line profiles at four different times following plasmoid instability are shown in Figure 9. These spectral line profiles are calculated as:
\begin{equation}
I\left(\lambda \right)= \frac{\int_{-0.02L_0}^{0.02L_0} \int_{0}^{L_0}\phi_{\lambda}n_en_HG\left(n_{ne},T\right)dx dy}{0.04L_0}
\end{equation}
where $n_{e}$ is the electron density, $n_H$ is the number density of hydrogen protons, $G\left( n_{e},T \right)$ is the contribution function that can be deduced from CHIANTI, and $\phi_{\lambda}$ is the relative velocity distribution function  and is given by:

\begin{equation}
\phi_{\lambda} =\frac{1}{\pi^{1/2} \Delta \lambda_{\rm D}} \exp \left[ -\left( \frac{\Delta \lambda +\lambda_0 \frac{v_{\rm x}}{c}}{\Delta \lambda_{\rm D}}  \right)^2 \right],
\end{equation}
where $\Delta \lambda=\lambda-\lambda_0$ is the offset from the rest wavelength $\lambda_0=1393.755$\AA\  and  $v_x$ is the velocity in the $x$-direction that is the line of sight direction. The thermal broadening of the line profile $\Delta \lambda_D$ is given by:
\begin{equation}
\Delta \lambda_D = \frac{\lambda_0}{c}\sqrt{\frac{2k_BT}{m_{Si}}},
\end{equation}
where $m_{Si}$ and $c$ represent the atomic mass of silicon and the speed of light, respectively.
 
As shown in Figure 9, the Si~{\sc{iv}} emission intensity gets stronger, and the line profile gets broader following the occurrence of the plasmoid instability. The emission intensity increases by more than four orders of magnitude in less than one second. At $t=5.74$\,s, the Si~{\sc{iv}} emission intensity reaches more than $10^7$\,erg\,cm$^{-2}$\,s$^{-1}$\,sr$^{-1}$\AA\ $^{-1}$, and gets close to the observed ones \citep{Young2018}. The width of the Si~{\sc{iv}} spectral line profile including multi-peaks reaches equivalently about $100$\,km\,s$^{-1}$, which is consistent with observations \citep[e.g.,][]{Peter2014, Tian2016}. Basically, the large width of the spectral line profile could be ascribed to the strong magnetic field or the low plasma density in the reconnection region. 

We ought to point out that these calculations are based on an optically thin assumption. In the future, it is worth looking into detailed properties of the width and the shape of the spectral line profile, as well as the emission intensity in the given wavelength by solving the radiative transfer equations.

\begin{figure*}
    \centerline{\includegraphics[width=0.45\textwidth, clip=]{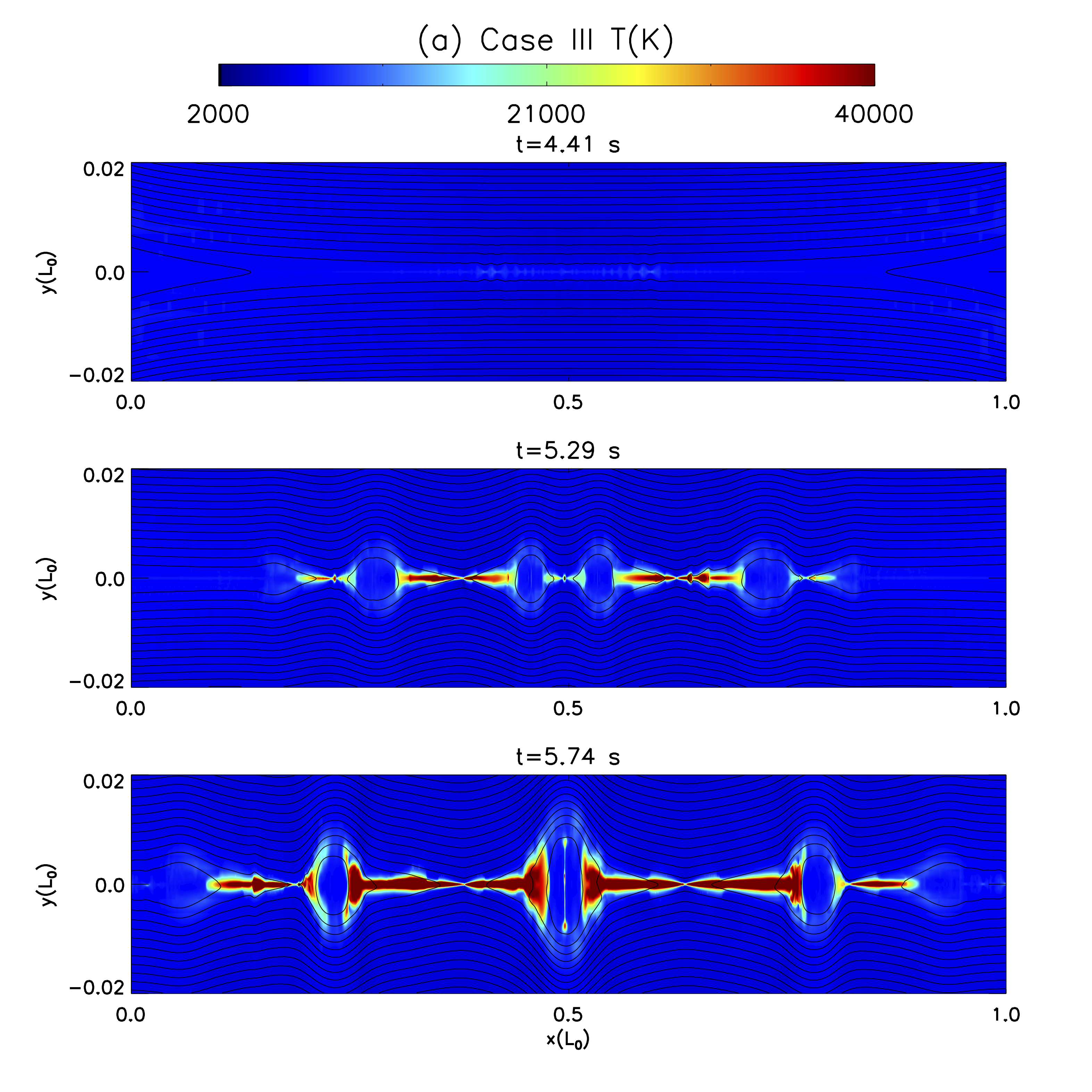}
                       \includegraphics[width=0.45\textwidth, clip=]{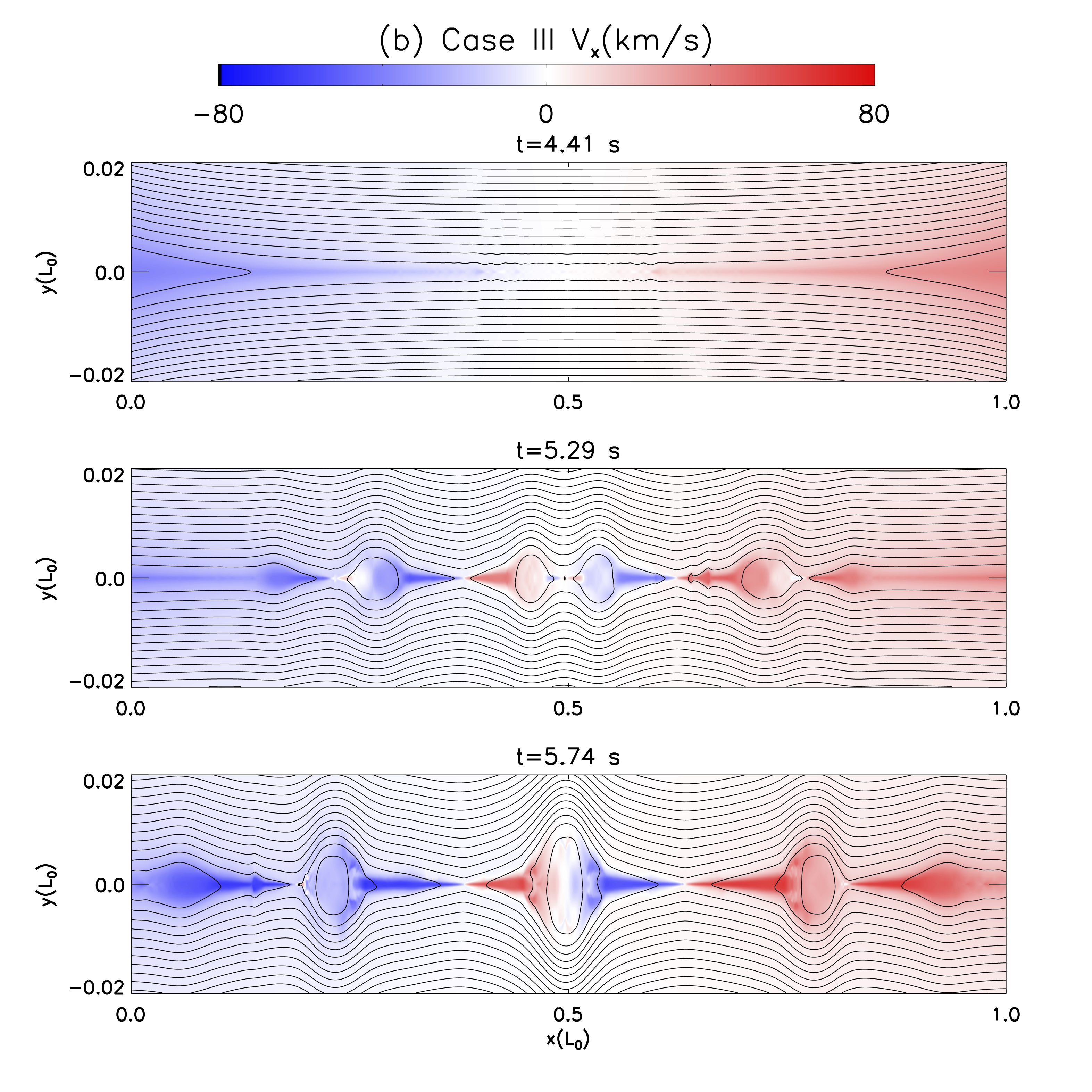}} 
  \caption{Distributions of the temperature (a), and the velocity in the $x$-direction ($v_x$) (b) at three different times in case III. The black solid lines represent the magnetic field lines, and the background colors in (a) and (b) represent the temperature $T$ and the velocity $v_x$ respectively.} \label{f8}
\end{figure*}

\begin{figure*}
    \centerline{\includegraphics[width=0.45\textwidth, clip=]{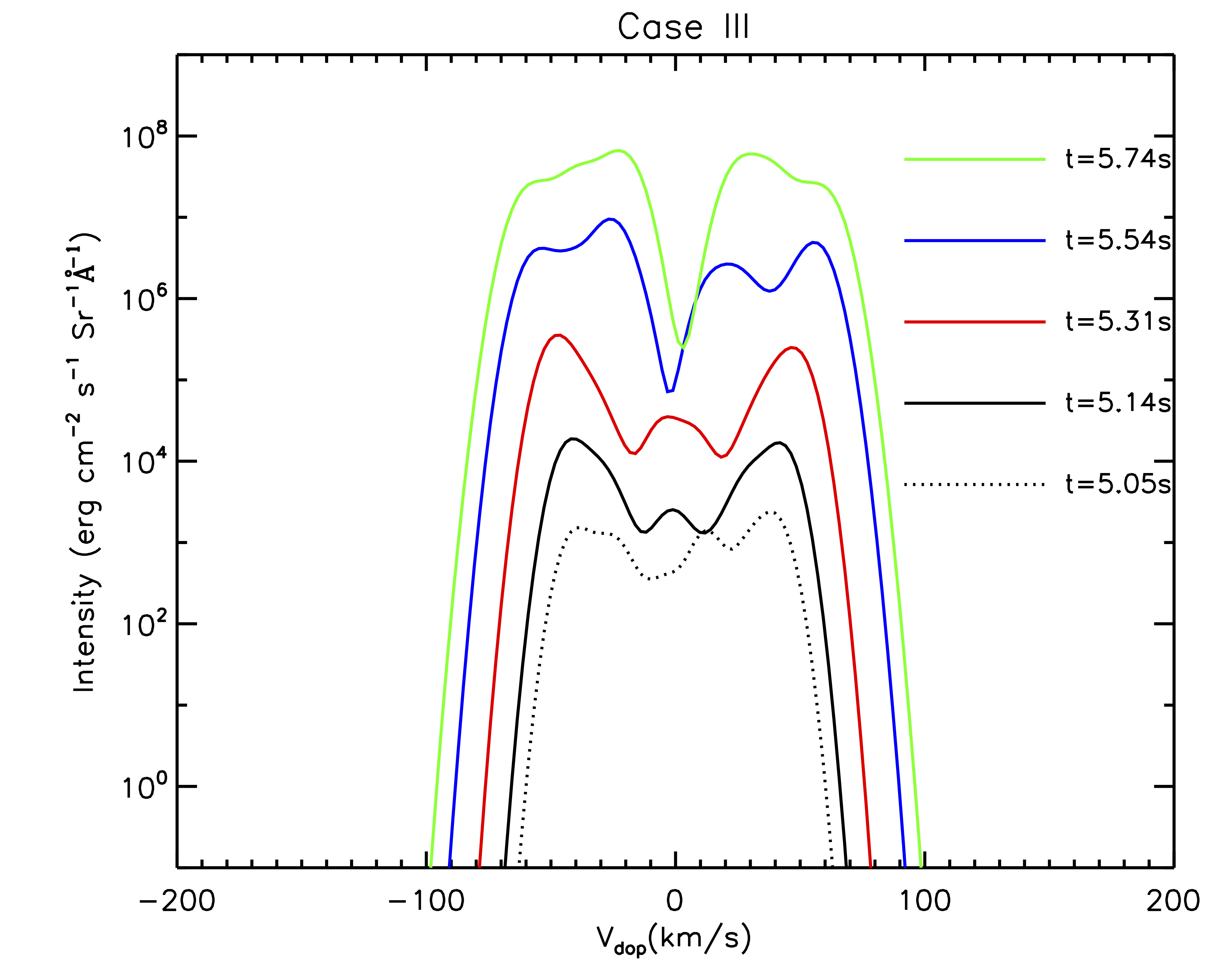}} 
  \caption{Synthesized Si IV spectral line profiles at five different times after the plasmoid instability appears in Case III. } \label{f9}
\end{figure*}

\subsection{Results with a much lower resolution} \label{sec:spec}
Numerical diffusion was used to trigger reconnection in previous RMHD simulations in 3D due to the limited grid resolution. In order to compare the difference between the results with more realistic diffusivities and the results triggered by numerical diffusion, we run Case IV with a resolution about 500 times lower than that in Case III. The low resolution in Case IV yields the numerical diffusion instead of the physical diffusions trigger magnetic reconnection. Figure 10 shows distributions of temperature and velocity in the $x$-direction. In comparing Figures 8 and 10, we can find that the large numerical diffusion in Case IV creates a much thicker current sheet, and leads to a later appearance of the plasmoid instability than what occurs in Case III. 

The plasma with a high temperature above $40,000$\,K starts to appear after the plasmoid instability takes place in Case III, but the temperature above $40,000$\,K occurs prior to the plasmoid instability in Case IV. The maximum outflow velocity in Case IV is lower compared to that in Case III. 
In addition, we also repeated the experiment in Case I with a much lower  grid resolution, but the result indicates that no apparent heating occurs in this low resolution simulation. 

These low-resolution tests imply that invoking reconnection in the current sheet by enhancing the numerical diffusion might yield an unexpected side-effect, which prevents important physical information of the studied phenomenon from being revealed properly. For the specific UV burst investigated here, the important physical information may include the mechanisms for the burst, emission intensity, and the profile of the given spectral lines.

\begin{figure*}
    \centerline{\includegraphics[width=0.45\textwidth, clip=]{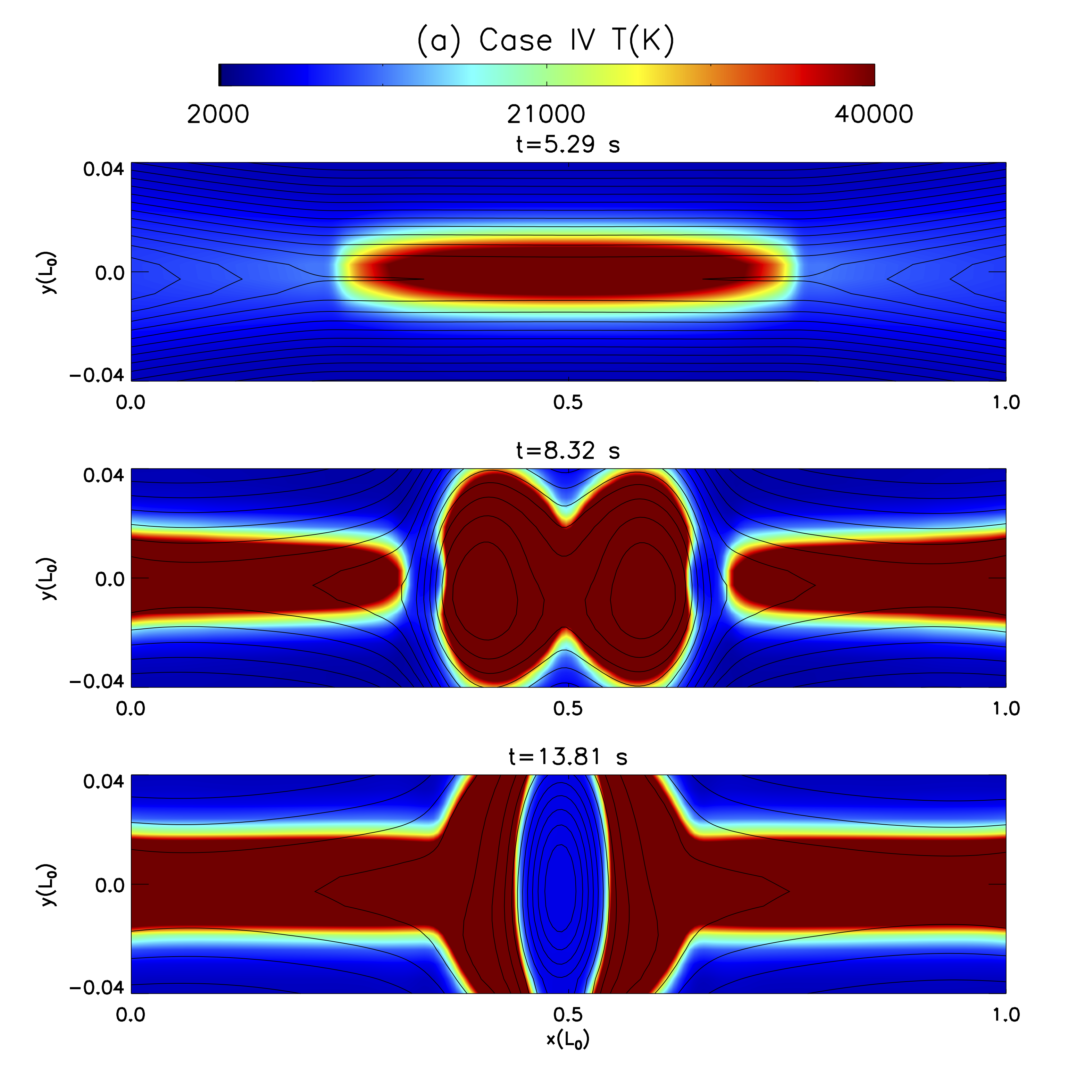}
                       \includegraphics[width=0.45\textwidth, clip=]{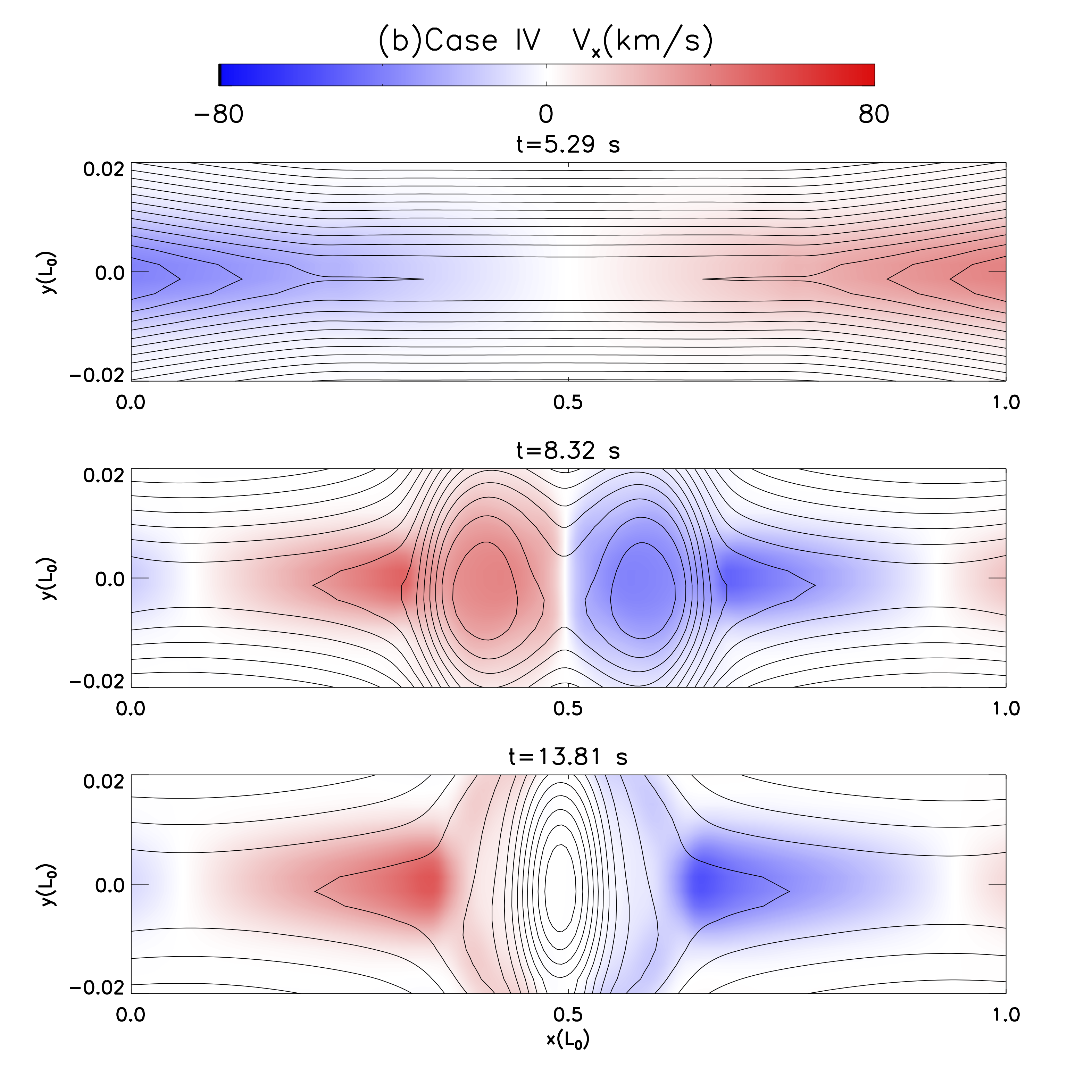}} 
  \caption{Distributions of the temperature (a), and the velocity in the $x$-direction ($v_x$) (b) at three different times in case IV. The black solid lines represent the magnetic field lines, and the background colors in (a) and (b) represent the temperature, $T$, and the velocity, $v_x$, respectively.} \label{f10}
\end{figure*}

%--------------------------------------------------------------------------------------------------------------------------%

\section{Summary and outlook}
In this work, we diligently investigate the low $\beta$ magnetic reconnection around the solar TMR. Compared with our previous works \cite{Ni2015,Ni2016, Ni2021}, the temperature-dependent ionization fractions of hydrogen and helium provide more realistic radiative cooling and more accurate diffusive coefficients in simulations of this work. We aim to find out if UV bursts can be generated in the low chromosphere around TMR, which is related to the question of whether UV bursts and EBs can be generated within the same atmospheric layer. We also investigated the energy conversion process during magnetic reconnection and the dominant mechanism that converts magnetic energy into heat. 

We studied the same low $\beta$ magnetic reconnection process with two different radiative cooling models, the results show that radiative cooling does not significantly change the reconnection mechanisms and the energy dissipation mechanisms in such a reconnection process with strong guide fields around TMR. The different radiative cooling models mostly affect the distributions of the temperature and density in an apparent fashion after plasmoid instability takes place. The radiative cooling model from \cite{Carlsson2012} is stronger than that from \cite{Gan1990}, and fewer plasmas are heated up to high temperatures when we use the \cite{Carlsson2012} model. However, the plasma with a high temperature $\sim20,000$\,K appears in both cases when the reconnection magnetic fields are stronger than $500$\,G, which is consistent with previous single-fluid and two-fluid simulation results \citep{Ni2015, Ni2016, Ni2018a, Ni2018b}. Since most of the neutral particles become ionized when the plasma is heated up to a high temperature above $20,000$\,K, the magnetic diffusion caused by electron-neutral collision ($\eta_{en}$) and ambipolar diffusion ($\eta_{AD}$) in these regions then decrease to a value that is much smaller than $\eta_{ei}$. Our main conclusions are given as follows:

1.   The UV burst can be generated in the low chromosphere with a high plasma density $\sim 10^{20}-10^{21}$\,m$^{-3}$ as long as the reconnection magnetic field is strong enough.  When the reconnection magnetic field is stronger than $900$\,G, the width of the synthesized Si IV 1394 {\AA} line profile with multiple peaks reaches up to $100$\,km\,s$^{-1}$, which is consistent with observations.

2. Joule heating which directly converts magnetic energy into thermal energy in the low $\beta$ reconnection process is mainly contributed by magnetic diffusion due to electron-ion collision ($\eta_{ei}$). Part of the kinetic energy generated by magnetic reconnection is also subsequently converted into thermal energy by local compression. The compression heating is dominant for producing thermal energy, which is much stronger than Joule heating after the turbulent reconnection mediated by plasmoids takes place.

3. The average power density of the generated thermal energy in the low $\beta$ reconnection region can reach above $1000$\,erg\,cm$^{-3}$\,s$^{-1}$, which is comparable to the average power density accounting for a UV burst.

The \cite{Carlsson2012} model is the most acceptable simple radiative cooling model for the chromosphere. However, such a model is more suitable for the middle and up chromosphere. In the future work, we might still need to solve the full radiation MHD equations with high resolutions to get more accurate results about the temperature evolutions and distributions in the reconnection region.  Since the low solar atmosphere is strongly stratified, the reconnection mechanisms above the middle chromosphere might be very different from the low chromosphere and the kinetic effects might start to become important \citep{Jara-Almonte2019, Jara-Almonte2021} when the plasma is not as dense as that in the low chromosphere. The UV bursts extending from the low chromosphere to the transition region include different reconnection mechanisms, energy dissipation mechanisms, and radiative transfer and cooling processes at different atmospheric layers, requiring additional  exploration in future studies.     	

\begin{acknowledgements}
We thank the referee for all the comments and suggestions. Lei Ni would like to thank professor Leenaarts and professor Carlsson for helpful discussions and suggestions. This research is supported by the NSFC Grants 11973083 and 11933009; the Strategic Priority Research Program of CAS with grants XDA17040507; the outstanding member of the Youth Innovation Promotion Association CAS (No. Y2021024 ); the Applied Basic Research of Yunnan Province in China Grant 2018FB009; the CAS Key Laboratory of Solar activity (Grant KLSA202103); the Yunling Talent Project for the Youth; the project of the Group for Innovation of Yunnan Province grant 2018HC023; the Yunling Scholar Project of the Yunnan Province and the Yunnan Province Scientist Workshop of Solar Physics; Yunnan Key Laboratory of Solar Physics and Space Science (No. 202205AG070009); the Special Program for Applied Research on Super Computation of the NSFC-Guangdong Joint Fund (nsfc2015-460, nsfc2015-463, the second phase); the data analysis is performed on the cluster in the Computational Solar Physics Laboratory of Yunnan Observatories.
\end{acknowledgements}


\end{document}